\documentclass[sigconf]{acmart}

\usepackage{booktabs} 

\usepackage{booktabs} 

\usepackage{fancyhdr}
\usepackage[normalem]{ulem}
\usepackage{microtype}

\usepackage{graphicx}
\usepackage{subfigure}
\usepackage{epstopdf}
\usepackage{url}
\usepackage{colortbl}
\definecolor{mygray}{gray}{.88}
\usepackage{multirow}
\usepackage{amsmath}
\usepackage{marvosym}
\usepackage{bm}

\newcommand{\tabincell}[2]{\begin{tabular}{@{}#1@{}}#2\end{tabular}}

\copyrightyear{2018}
\acmYear{2018}
\setcopyright{acmcopyright}
\acmConference[PACT '18]{International conference on Parallel Architectures and Compilation Techniques}{November 1--4, 2018}{Limassol, Cyprus} \acmBooktitle{International conference on Parallel Architectures and Compilation Techniques (PACT '18), November 1--4, 2018, Limassol, Cyprus} \acmPrice{15.00}
\acmDOI{10.1145/3243176.3243190} \acmISBN{978-1-4503-5986-3/18/11}

\begin{document}
\title{Data Motifs: A Lens Towards Fully Understanding Big Data and AI Workloads}

\author{Wanling Gao}
\affiliation{%
  \institution{State Key Laboratory of Computer Architecture\\Institute of Computing Technology, Chinese Academy of Sciences\\University of Chinese Academy of Sciences}
}
\email{gaowanling@ict.ac.cn}

\author{Jianfeng Zhan}
\authornote{Jianfeng Zhan is the corresponding author.}
\affiliation{%
  \institution{State Key Laboratory of Computer Architecture\\Institute of Computing Technology, Chinese Academy of Sciences\\University of Chinese Academy of Sciences}
}
\email{zhanjianfeng@ict.ac.cn}

\author{Lei Wang}
\affiliation{%
  \institution{State Key Laboratory of Computer Architecture\\Institute of Computing Technology, Chinese Academy of Sciences}
}
\email{wanglei_2011@ict.ac.cn}

\author{Chunjie Luo}
\affiliation{%
  \institution{Institute of Computing Technology, Chinese Academy of Sciences}
}
\email{luochunjie@ict.ac.cn}

\author{Daoyi Zheng}
\affiliation{%
  \institution{Institute of Computing Technology, Chinese Academy of Sciences}
}
\email{zhengdaoyi@baidu.com}

\author{Fei Tang}
\affiliation{%
  \institution{Institute of Computing Technology, Chinese Academy of Sciences}
}
\email{tangfei@ict.ac.cn}

\author{Biwei Xie}
\affiliation{%
  \institution{Institute of Computing Technology, Chinese Academy of Sciences}
}
\email{xiebiwei@ict.ac.cn}

\author{Chen Zheng}
\affiliation{%
  \institution{Institute of Computing Technology, Chinese Academy of Sciences}
}
\email{zhengchen@ict.ac.cn}

\author{Xu Wen}
\affiliation{%
  \institution{University of Chinese Academy of Sciences}
}
\email{wenxu@ict.ac.cn}

\author{Xiwen He}
\affiliation{%
  \institution{Institute of Computing Technology, Chinese Academy of Sciences}
}
\email{hexiwen@ict.ac.cn}

\author{Hainan Ye}
\affiliation{%
  \institution{Beijing Academy of Frontier Sciences and Technology}
}
\email{yehainan@mail.bafst.com}

\author{Rui Ren}
\affiliation{%
  \institution{Institute of Computing Technology, Chinese Academy of Sciences}
}
\email{renrui@ict.ac.cn}

\begin{abstract}

The complexity and diversity of big data and AI workloads make understanding them difficult and challenging.
This paper proposes a new approach to modelling and characterizing big data and AI workloads. We consider each big data and AI workload as a pipeline of one or more classes of units of computation performed on different initial or intermediate data inputs. Each class of unit of computation captures the common requirements  while being reasonably divorced from individual implementations, and hence we call it a \emph{data motif}.
For the first time, among a wide variety of big data and AI workloads, we identify eight \emph{data motifs} that  take up most of the run time of those workloads, including \emph{Matrix}, \emph{Sampling}, \emph{Logic}, \emph{Transform},
\emph{Set}, \emph{Graph}, \emph{Sort} and \emph{Statistic}.
We implement the eight data motifs on different software stacks as the micro benchmarks of an open-source big data and AI benchmark suite --- BigDataBench 4.0 (publicly available from \url{http://prof.ict.ac.cn/BigDataBench}), and perform comprehensive characterization of those data motifs from perspective of data sizes, types, sources, and patterns as a lens towards fully understanding big data and AI workloads. We believe the eight data motifs are promising abstractions and tools for not only big data and AI benchmarking, but also domain-specific hardware and software co-design.

\end{abstract}

%
%



\maketitle
\thispagestyle{firstpage}
\pagestyle{plain}

\section{Introduction}

The complexity and diversity of big data and AI workloads make understanding them difficult and challenging.
First, modern big data and AI workloads expand and change very fast, and it is impossible
to create a new benchmark or proxy for every possible
workload.
Second, several fundamental changes, i.e., end of Dennard scaling, ending of Moore's Law, Amdahl's Law and its implications for ending "Easy" multicore era, indicate only hardware-centric path left is Domain-specific Architectures~\cite{2018_turing}. To achieve higher efficiency, we need tailor the architecture to
characteristics of a domain of applications~\cite{2018_turing}. However, the first step is to understand Big Data and AI workloads. 
Third, whatever early in the architecture design process
or later in the system evaluation, it is time-consuming to run
a comprehensive benchmark suite. The complex software
stacks of the modern workloads aggravate this issue. The modern big data or AI benchmark suites~\cite{wang2014bigdatabench, ferdman2011clearing} are too huge to run on simulators and hence challenge time-constrained simulation and even make it impossible.
Fourth, too complex workloads raise challenges in both reproducibility
and interpretability of performance data in benchmarking systems.

Identifying abstractions of time-consuming units of computation
is an important step toward fully understanding complex workloads.
Much previous work~\cite{codd1970relational,chen2014tpc,colella2004defining,asanovic2006landscape,shah2010data} has illustrated the importance of abstracting workloads in corresponding domains. TPC-C~\cite{chen2014tpc} is a successful benchmark built on the basis of frequently-appearing operations in the OLTP domain.
HPCC~\cite{luszczek2006hpc} adopts a similar method to design a benchmark suite for high performance computing.
National Research Council proposes seven major tasks in massive data analysis~\cite{council2013frontiers}, while they are macroscopical definition of problems from the perspective of mathematics.
Unfortunately, to the best of our knowledge, none of previous work has identified time-consuming classes of unit of computation in big data and AI workloads.  

Also, identifying abstractions of time-consuming units of computation is an important step toward domain-specific hardware and software co-design. Straightforwardly,  we can tailor the architecture to characteristics of an application, several applications, or even a domain of applications~\cite{2018_turing}. The past witnesses the success of  neural network processors for machine learning~\cite{jouppi2017datacenter, chen2014diannao}, GPUs for graphics, virtual reality~\cite{owens2008gpu}, and programmable network switches and interfaces~\cite{2018_turing}. Moreover, if we can identify abstractions of time-consuming units of computation in Big Data and AI workloads and design domain-specific hardware and software system for them, our target will be much general-purpose. Meanwhile, optimizing most time-consuming units of computation other than many algorithms case by case on different hardware or software systems will be much efficient.

In this paper, we propose a new approach to modelling and characterizing big data and AI workloads. We consider each big data and AI workload as a pipeline of one or more classes of unit of computation performed on different initial or intermediate data inputs, each
of which captures the common requirements while being
reasonably divorced from individual implementations~\cite{asanovic2006landscape}. We call this abstraction \emph{a data motif}. \emph{Significantly different from the traditional kernels, a data motif's behaviors are affected by the sizes, patterns, types,  and sources of different data inputs; Moreover, it reflects not only computation patterns, memory access patterns, but also disk and network I/O patterns}.

After thoroughly analyzing a majority of workloads in five
typical big data application domains (search engine, social network, e-commerce, multimedia and bioinformatics), we identify eight data motifs that take up most of run time,
including \emph{Matrix}, \emph{Sampling}, \emph{Logic}, \emph{Transform},
\emph{Set}, \emph{Graph}, \emph{Sort} and \emph{Statistic}. We found the combinations of one or more data motifs with different weights in terms of runtime can
describe most of big data and AI workloads we investigated~\cite{gao2018proxy}.
Considering various data inputs---text, sequence, graph, matrix and image data---with different data types and distributions,  we implement eight data motifs on different software stacks, including Hadoop~\cite{hadoopweb}, Spark~\cite{zaharia2010spark}, TensorFlow~\cite{abadi2016tensorflow} and POSIX-thread (Pthread)~\cite{barney2009posix}.  For big data, the implemented data motifs include sort (\emph{Sort}), wordcount (\emph{Statistics}), grep (\emph{Set}), MD5 hash (\emph{Logic}), matrix multiplication (\emph{Matrix}), random sampling (\emph{Sampling}), graph traversal (\emph{Graph}) and FFT transformation (\emph{Transform}), while for AI, we implement 2-dimensional convolution (\emph{Transform}), max pooling (\emph{Sampling}), average pooling (\emph{Sampling}), ReLU activation (\emph{Logic}), sigmoid activation (\emph{Matrix}), tanh activation (\emph{Matrix}), fully connected (\emph{Matrix}), and element-wise multiplication (\emph{Matrix}), which are frequently-used computation in neural network modelling. We release the implemented data motifs as the micro benchmarks of an open-source big data and AI benchmark suite --- BigDataBench. In the rest of paper, we use the big data motifs to indicate the motif implementations for big data, and use the AI motifs to indicate the motif implementations for AI.

Just like relation
algebra in database,  the data  motifs are promising fundamental
concepts and tools for benchmarking, designing, measuring,
and optimizing big data and AI systems.
Based on the data motifs, we build the fourth version of BigDataBench~\cite{gao2018bigdatabench}, including micro benchmarks, each of which is a data motif, and component benchmarks, each of which is a combination of several data motifs, and end-to-end application benchmarks, each of which is a combination of component benchmarks. Also, we build the proxy benchmarks~\cite{gao2018proxy} for big data and AI workloads, which has a speedup up to 1000 times in terms of runtime and a micro-architectural data accuracy of more than 90\%.
In this paper, as the first step, we call attention to  performing  comprehensive characterization of those data motifs from perspective of data sizes, types, sources, and patterns as a lens towards fully understanding big data and AI workloads.
On a typical state-of-practice processor: Intel Xeon E5-2620 V3, we comprehensively characterize all data motif implementations and identify their bottlenecks.

Our contributions are five-fold as follows:

\begin{itemize}
\item We identify eight data motifs through  profiling  a wide variety of big data and AI workloads.
\item We provide diverse data motif implementations  on the software stacks of Hadoop, Spark, TensorFlow, Pthread.
\item From the system and micro-architecture perspectives, we comprehensively characterize the behaviors of data motifs and identify their bottlenecks.
We find that these data motifs cover a wide variety of performance space, from the perspectives of system and micro-architecture behaviors.
Moreover, the behavior of each motif is not only influenced by its algorithm, but also largely affected by the type, source, size, and pattern of input data.
\item From the system aspect,
we find that some AI motifs like convolution, fully-connected are CPU-intensive, while the other AI motifs are not CPU-intensive, such as Relu, Sigmoid used as activation layer.
Further, the AI motifs have little pressure on disk I/O, since they load a batch (e.g. 128 images) from disk every iteration.
    \item From the micro-architecture aspect, we find that these motifs show  various computation and memory access patterns, exploiting different parallelism degrees of ILP and MLP. With the data size expanding, the percentage of frontend bound decreases while the backend bound increases.

\end{itemize}

The rest of the paper is organized as follows.
Section 2 illustrates the motivation of identifying data motifs.
Section 3 introduces data motif identification methodology. Section 4 performs system and micro-architecture evaluations on the data motif implementations. In Section 5, we report the data impact on the data motifs' behaviors from perspectives of data size, data pattern, data type and data source.
Section 6 introduces the related work. Finally, we draw a conclusion in Section 7.

\section{Motivation}

We take two examples to explain why we should call attention to performing comprehensive characterization of those data motifs.

\subsection{\textbf{SIFT Workload in Computer Vision}}

\begin{figure}[!t]
\centering
\includegraphics*[scale=0.37]{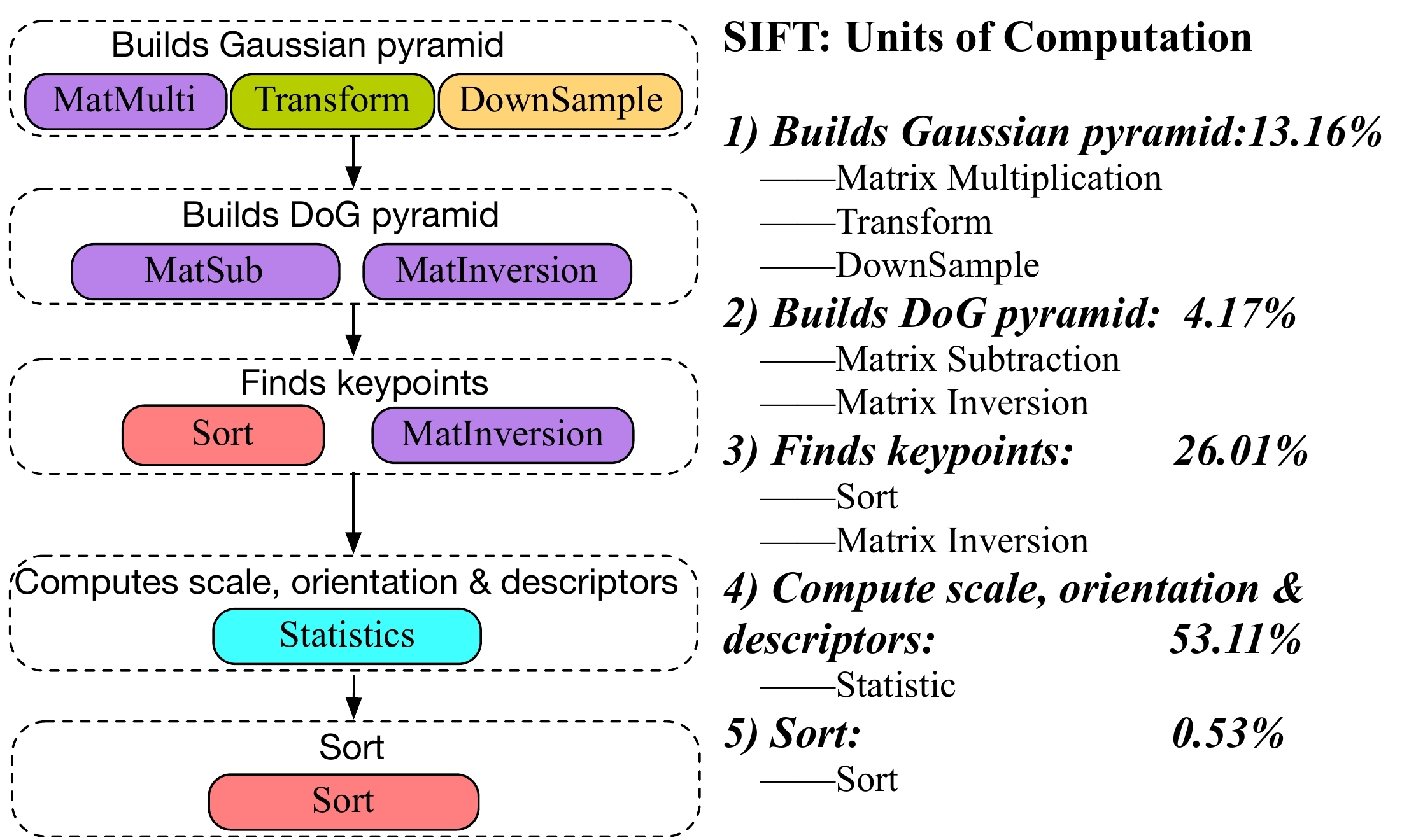}
\caption{The Computation Dependency Graph and Run Time Breakdown of SIFT Workload.} 
\label{SIFT}
\end{figure}

SIFT~\cite{lowe2004distinctive} is a typical workload for feature extraction, and widely used to detect local features of input images.

Fig.~\ref{SIFT} shows the computation dependency graph and run time breakdown of SIFT workload. 
In total, SIFT involves five data motifs.
Gaussian filters $G(x,y,\partial)$ with different space scale factors $\partial$ are used to generate a group of image scale spaces, through the convolution with the input image.
Image pyramid is to downsample these image scale spaces.
DOG image means difference-of-Gaussian image, which is produced by matrix subtraction of adjacent image scale spaces in image pyramid.
After that, every point in one DOG scale space would sort with eight adjacent points in the same scale space and points in adjacent two scale spaces, to find the key points in the image.
Through profiling, we find that \emph{computes descirptors, finds keypoints} and \emph{builds gaussian pyramid} are three main time-consuming parts of  the SIFT workload. Furthermore, we analyze those three parts and find they consist of  several classes of unit of computation, like Matrix, Sampling, Transform, Sort and Statistics, summing up to 83.23\% of the total SIFT run time. 

\subsection{\textbf{AlexNet in AI}}

\begin{figure}[!t]
\centering
\includegraphics*[scale=0.37]{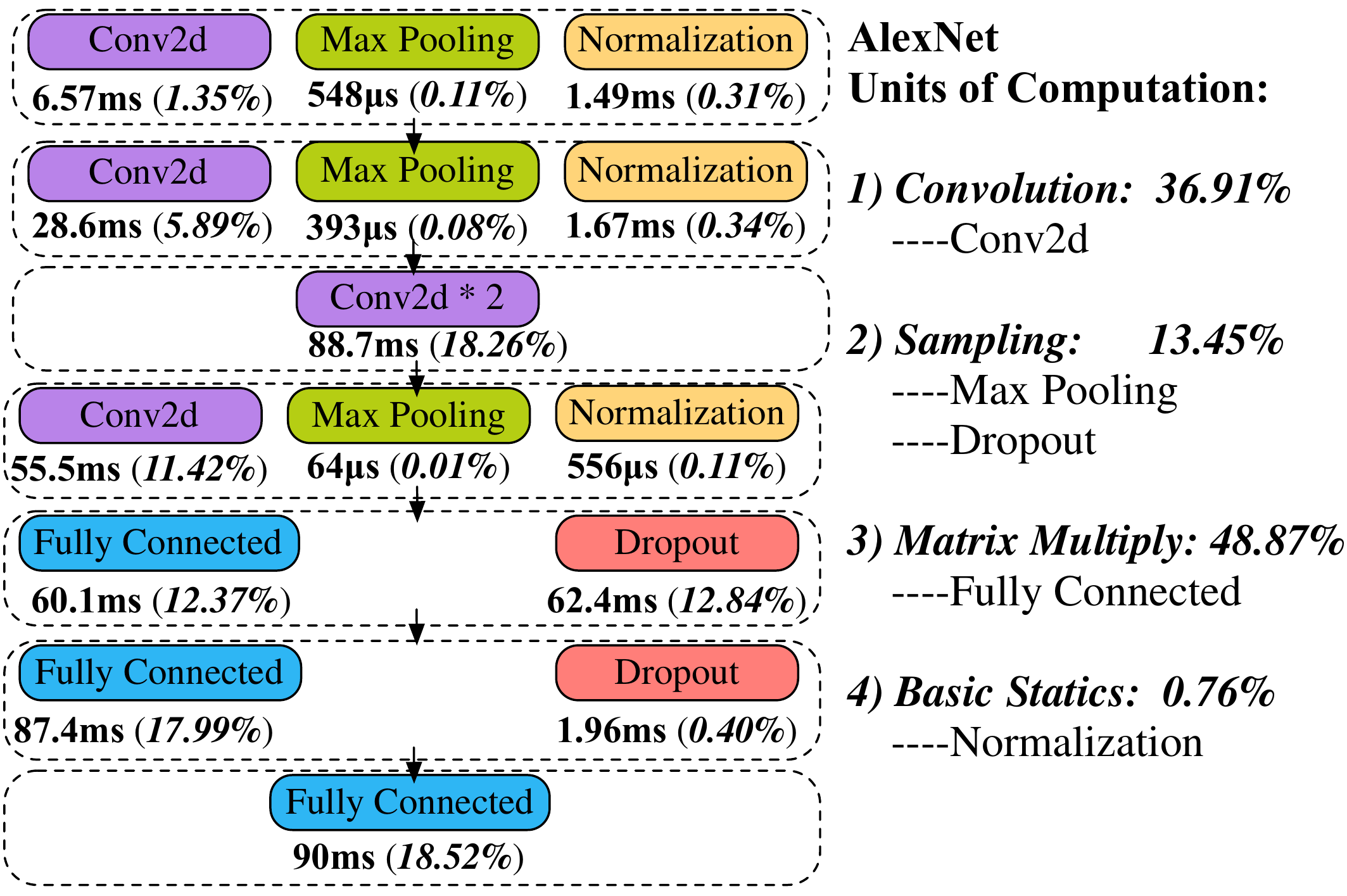}
\caption{The Computation Dependency Graph and Run Time Breakdown of One Iteration of TensorFlow AlexNet Workload.} 
\label{alexnet}
\end{figure}

AlexNet~\cite{krizhevsky2012imagenet} is a representative and widely-used convolutional neural network in deep learning.
In total, it has eight layers, including five convolutional layers and three fully connected layers.

We profile one iteration of the AlexNet workload (implemented with TensorFlow) using TensorBoard toolkit.  Fig.~\ref{alexnet} presents its  computation dependency graph and run time breakdown. For each operator, we report its run time and its percentage of the total run time, such as 6.57 ms and 1.35\% for the first convolution operator.
We find that each iteration involves Transform (conv2d), Sampling (max pooling, dropout),  Statistics (normalization), and Matrix (fully connected). Among them, matrix and transform computations occupy a large proportion---48.87\% and 36.91\%, respectively.

Through the above analysis, we have the following observation.
Though big data and AI workloads are very complex and fast-changing, we can consider them as a pipeline of one or more fundamental classes of unit of computation performed on different initial or intermediate data inputs.
Those classes of unit of computation, which we call data motifs, occupy most of the run time of the workloads, so we should pay more attention to them.
In the next section, we will investigate more extensive big data and AI workloads, and elaborate the design of data motifs.

\section{Methodology}

Data motifs are frequently-appearing classes of unit of computation handling different data inputs.
In this section, we illustrate how to identify data motifs from big data and AI workloads, and illustrate our data motif implementations.

\begin{table*}[htb]
\caption{The Importance of Eight Data motifs in Big Data and AI workloads.}\label{statistical}
\renewcommand\arraystretch{1.2}
\small
\center
\begin{tabular}{|p{1in}|p{1.2in}|p{2.1in}|p{2in}|}
  \hline
  \textbf{Catergory}   & \textbf{ Application Domain } &  \textbf{Workload } &  \textbf{ Unit of Computation } \\
  \hline
  \multirow{2}{*}{\tabincell{l}{Deep Learning}} &  \multirow{2}{*}{\tabincell{l}{ Image Recognition \\ Speech Recognition }} & Convolutional neural network(CNN)  &  Matrix, Sampling, Transform \\  \cline{3-4}
                                                &  & Deep belief network(DBN)         & Matrix, Sampling \\ \cline{3-4}
  \hline

  \multirow{2}{*}{\tabincell{l}{Graph Mining}}  &  \multirow{2}{*}{\tabincell{l}{ Search Engine \\ Community Detection }}  & PageRank  & Matrix, Graph, Sort \\  \cline{3-4}
                                                &   & BFS, Connected component(CC)      &  Graph \\ \cline{3-4}
  \hline

  \multirow{2}{*}{\tabincell{l}{Dimension Reduction}}  &  \multirow{2}{*}{\tabincell{l}{ Image Processing \\ Text Processing }}  & Principal components analysis(PCA)  & Matrix \\  \cline{3-4}
                                                &   & Latent dirichlet allocation(LDA)  &  Statistics, Sampling \\ \cline{3-4}
  \hline

  \multirow{3}{*}{\tabincell{l}{Recommendation}} & \multirow{3}{*}{\tabincell{l}{Association Rules Mining \\ Electronic Commerce }}  & Aporiori  &  Statistics, Set \\  \cline{3-4}
                                                             &  & FP-Growth &  Graph, Set, Statistics \\ \cline{3-4}
                                                             &  & Collaborative filtering(CF) &  Graph, Matrix \\ \cline{3-4}
  \hline

  \multirow{4}{*}{\tabincell{l}{Classification}} &  \multirow{4}{*}{\tabincell{l}{ Image Recognition \\ Speech Recognition \\ Text Recognition }} & Support vector machine(SVM)  &  Matrix \\  \cline{3-4}
                                                 &  & K-nearest neighbors(KNN) &  Matrix, Sort, Statistics \\ \cline{3-4}
                                                 &  & Naive bayes  &  Statistic \\ \cline{3-4}
                                                 &  & Random forest  &  Graph, Statistics  \\ \cline{3-4}
                                                 &  & Decision tree(C4.5/CART/ID3)  &  Graph, Statistics  \\ \cline{3-4}
  \hline

  \multirow{1}{*}{\tabincell{l}{Clustering}} & \multirow{1}{*}{\tabincell{l}{ Data Mining }} &  K-means  & Matrix, Sort  \\ \cline{3-4}
  \hline

  \multirow{4}{*}{\tabincell{l}{Feature Preprocess}} & \multirow{4}{*}{\tabincell{l}{ Image Processing \\ Signal Processing \\ Text Processing}}  & Image segmentation(GrabCut)  & Matrix, Graph \\  \cline{3-4}
                                                     &  & Scale-invariant feature transform(SIFT)  & Matrix, Transform, Sampling, Sort, Statistics \\  \cline{3-4}
                                                     &  & Image Transform  & Matrix, Transform \\  \cline{3-4}
                                                     &  & Term Frequency-inverse document frequency (TF-IDF)   & Statistics \\  \cline{3-4}
  \hline

  \multirow{2}{*}{\tabincell{l}{Sequence Tagging}} &\multirow{2}{*}{\tabincell{l}{ Bioinformatics \\ Language Processing }}  & Hidden Markov Model(HMM)  &  Matrix \\  \cline{3-4}
                                               &  & Conditional random fields(CRF) & Matrix, Sampling \\ \cline{3-4}
  \hline


  Indexing & Search Engine & Inverted index, Forward index & Statistics, Logic, Set, Sort\\
  \hline

  \multirow{4}{*}{\tabincell{l}{Encoding/Decoding}} & \multirow{4}{*}{\tabincell{l}{ Multimedia Processing \\ Security \\ Cryptography \\ Digital Signature }}  & MPEG-2  & Matrix, Transform \\  \cline{3-4}
                                                    &   & Encryption   & Matrix, Logic \\  \cline{3-4}
                                                    &   & SimHash, MinHash  & Set, Logic \\  \cline{3-4}
                                                    &  & Locality-sensitive hashing(LSH)  & Set, Logic \\ \cline{3-4}
  \hline

  \multirow{1}{*}{\tabincell{l}{Data Warehouse}} & \multirow{1}{*}{\tabincell{l}{  Business intelligence }}  &  Project, Filter, OrderBy, Union  &  Set, Sort \\  \cline{3-4}
  \hline


\end{tabular}
\end{table*}

\subsection{Motif Identification Methodology}

\begin{figure}[!t]
\centering
\includegraphics*[scale=0.8]{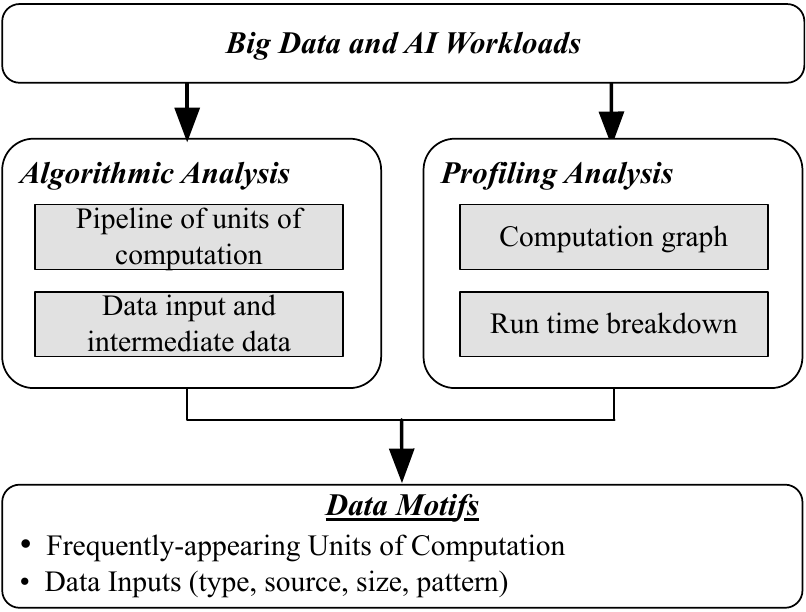}
\caption{Identifying Data Motifs.} 
\label{fig_motif_overview}
\end{figure}

Fig.~\ref{fig_motif_overview} overviews the methodology of motif identification.
We first single out a broad spectrum of big data and AI workloads through investigating five typical application domains (search engine, social network, e-commerce, multimedia, and bioinformatics) and representative algorithms in four processing techniques (machine learning, data mining, computer vision and natural language processing).
Then we conduct algorithmic analysis and profiling analysis on these workloads.
We profile the workload to analyze the computation dependency graph and run time breakdown, to find and correlate the hotspot functions to the code segments. Combing with algorithmic analysis, we decompose the workload into a pipeline of units of computation and focus on the input/intermediate data as well. Then we summarize the frequently-appearing and time-consuming units as data motifs. We repeat this procedure on forty workloads with a broad spectrum to guarantee the representativeness of our data motifs.

According to the units of computation pipeline and run time breakdown, we finalize eight big data and AI motifs, which are essential computations that take up most of run time.
Table~\ref{statistical} shows the importance of eight data motifs in a majority of big data and AI workloads.
Note that previous work~\cite{guinard2010resource} has identified four basic units of computation in online service, including get, put, post, delete. We don't include those four in our motif set. 


\subsection{Eight Data Motifs} \label{bigdatamotif}

In this subsection, we summarize eight data motifs that frequently appear in big data and AI workloads.

\textbf{Matrix} In big data and AI workloads, many problems involve matrix computations, such as vector-vector, matrix-vector and matrix-matrix operations.

\textbf{Sampling} 
Sampling plays an essential role in big data and AI processing, which selects a subset samples according to  certain statistical population. It can be used to obtain an approximate solution when one problem cannot be solved by deterministic method.

\textbf{Logic} We name computations performing bit manipulation as logic computations, such as hash, data compression and encryption.

\textbf{Transform} The transform computations here mean the conversion from the original domain (such as time) to another domain (such as frequency). Common transform computations include discrete fourier transform (DFT), discrete cosine transform (DCT) and wavelet transform.

\textbf{Set} In mathematics, Set means a collection of distinct objects. Likewise, the concept of Set is widely used in computer science.
Set is also the foundation of relational algebra~\cite{maier1983theory}. In addition,  similarity analysis of two data sets involves set computations, such as Jaccard similarity. Furthermore, fuzzy set and rough set play very important roles in computer science.

\textbf{Graph} A lot of applications involve graphs,
with nodes representing entities and edges representing dependencies.
Graph computation is notorious for having irregular memory access patterns.

\textbf{Sort} Sort is widely used in many areas. Jim Gray thought sort is the core of modern databases~\cite{asanovic2006landscape}, which shows its fundamentality.

\textbf{Statistics} Statistic computations are used to obtain the summary information through statistical computations, such as counting and probability statistics.

\subsection{Data Motif Implementations} \label{component}

Data motifs are the fundamental components of big data and AI workloads, which is of great significance for evaluation, considering the complexity and diversity of big data and AI workloads.
We provide the data motif implementations for big data and AI separately, according to their computation specialties.
For the big data motif implementations, we provide Hadoop~\cite{hadoopweb}, Spark~\cite{zaharia2010spark}, and Pthreads~\cite{barney2009posix} implementations. 
These data motifs include sort, wordcount, grep, MD5 hash, matrix multiplication, random sampling, graph traversal and FFT transformation.
For the AI motifs, we provide TensorFlow~\cite{abadi2016tensorflow} and Pthread implementations,
 including 2-dimensional convolution, max pooling, average pooling, relu activation, sigmoid activation, tanh activation, fully connected (matmul), and element-wise multiply.
We consider the impact of data input from the perspectives of type, source, size, and pattern. Among them, \emph{data type} includes structure, un-structured, and semi-structured data. \emph{Data source} indicates the data storage format, including text, sequence, graph, matrix, and image data. \emph{Data pattern} includes the data distribution, data sparsity, et al. As for \emph{data size}, we provide big data generators for text, sequence, graph and matrix data to fulfill different size requirements.

\section{Characterization}\label{evaluation}

In this section, we evaluate data motifs with various software stacks from the perspectives of both system and architecture behaviors.

\subsection{Experiment Setups}

We deploy a three-node cluster, with one master node and two slave nodes.
They are connected using 1Gb Ethernet network.
Each node is equipped with two Intel Xeon E5-2620 V3 (Haswell) processors, and each processor has six physical out-of-order cores.
The memory of each node is 64 GB.
The operating system, software stacks and gcc versions are as follows: CentOS 7.2 (with kernel 4.1.13); JDK 1.8.0\_65; Hadoop 2.7.1; Spark 1.5.2; TensorFlow 1.0; GCC 4.8.5.
The data motifs implemented with Pthread are compiled using "-O2" option for optimization.
The hardware and software details are listed in Table \ref{hwconfigeration}.
Since Pthread is a multi-thread programming model, we evaluate both the TensorFlow and Pthread implementations of AI motifs on one node for apple-to-apple comparison.

\begin{table}
\caption{Configuration Details of Xeon E5-2620 V3}\label{hwconfigeration}
\renewcommand\arraystretch{1.2}
\center
\footnotesize
\begin{tabular}{|p{0.64in}|p{0.64in}|p{0.64in}|p{0.64in}|}
\hline \rowcolor{mygray} \multicolumn{4}{|l|}{Hardware Configurations}\\
\hline \multicolumn{2}{|c|}{CPU Type} & \multicolumn{2}{c|}{Intel CPU Core} \\
\hline \multicolumn{2}{|c|}{Intel \textregistered Xeon E5-2620 V3}  &\multicolumn{2}{c|}{12 cores@2.40G} \\
\hline L1 DCache &L1 ICache &L2 Cache &L3 Cache \\
\hline 12 $\times$ 32 KB& 12 $\times$ 32 KB&12 $\times$ 256 KB& 15MB \\
\hline \multicolumn{2}{|c|}{Memory} & \multicolumn{2}{c|}{64GB,DDR4}  \\
\hline \multicolumn{2}{|c|}{Disk} & \multicolumn{2}{c|}{SATA@7200RPM}\\
\hline \multicolumn{2}{|c|}{Ethernet} & \multicolumn{2}{c|}{1Gb}\\
\hline \multicolumn{2}{|c|}{Hyper-Threading} & \multicolumn{2}{c|}{Disabled}\\
\hline
\end{tabular}
\end{table}

\subsection{Experiment Methodology}

\begin{figure*}[!t]
\centering
\includegraphics*[scale=0.7]{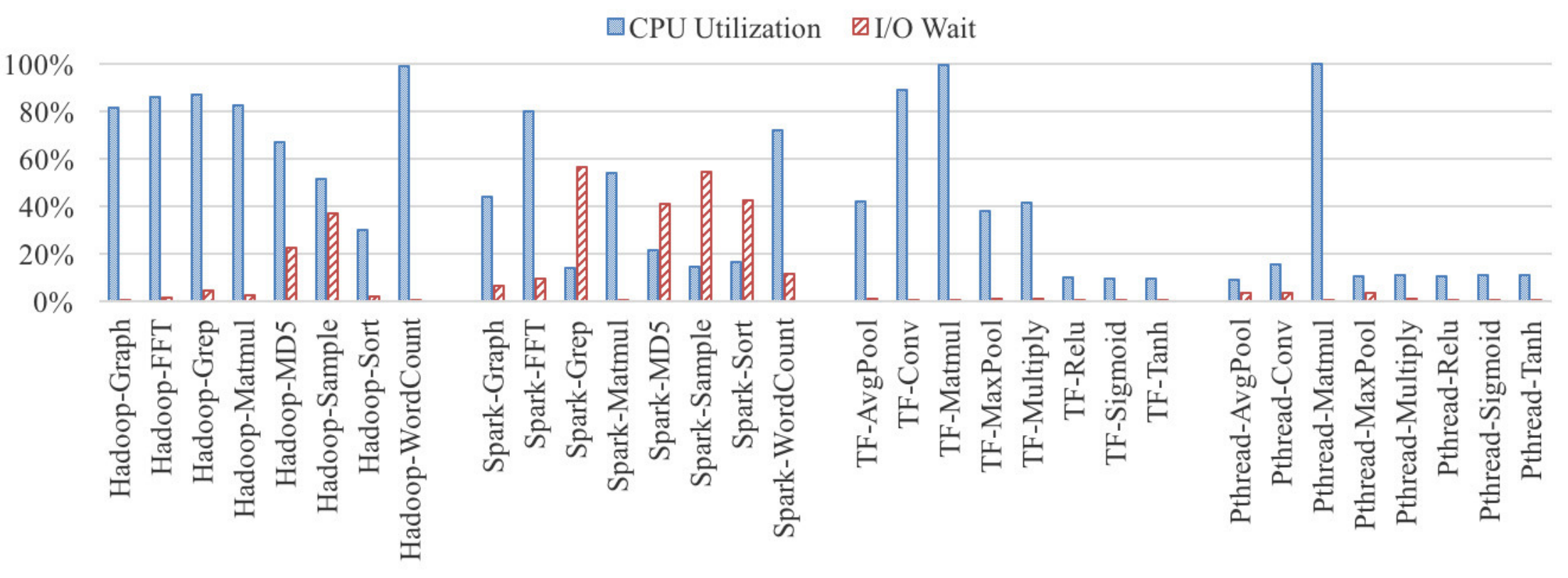}
\caption{CPU Utilization and I/O Wait of Data Motifs.} 
\label{cpuiowait}
\end{figure*}

\begin{figure}[!t]
\centering
\includegraphics*[scale=0.55]{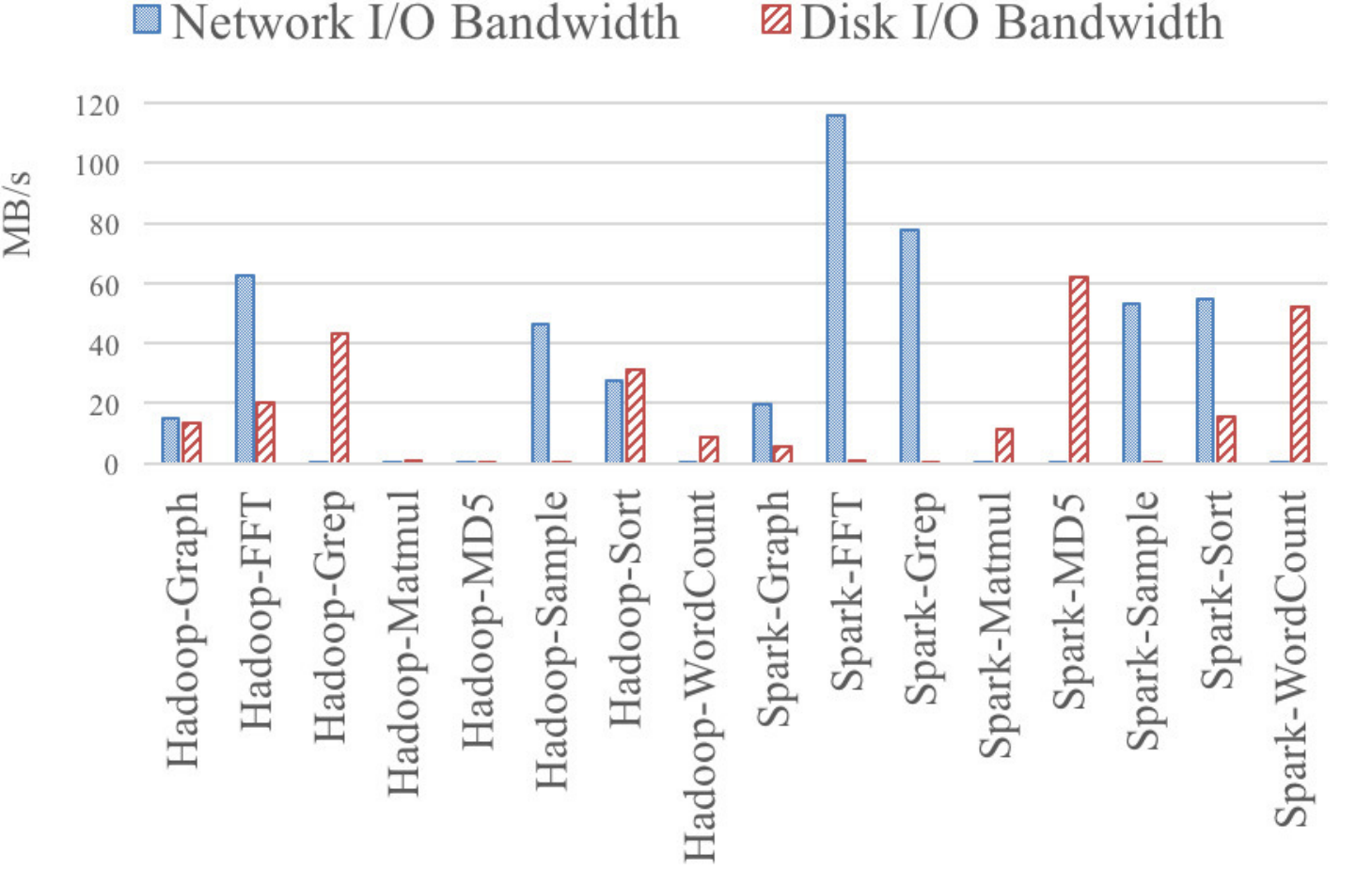}
\caption{I/O Behaviors of Data Motifs.} 
\label{io}
\end{figure}

We evaluate eight big data motifs implemented with Hadoop, Spark, and eight AI data motifs implemented with TensorFlow and Pthread.
Note that we use the optimal configurations for each software stack, according to the cluster scale and memory size.
The data configuration and selected metrics are listed as follows.

\textbf{Data Configuration}
To evaluate the impacts of data input comprehensively, we evaluate the data motifs with three data sizes: \emph{Small}, \emph{Medium}, and \emph{Large}. We choose the \emph{Large} data size according to the memory capacity of the cluster so as to fully utilize the memory resources, and the other two are chosen for comparison.
For the graph motif, \emph{Small}, \emph{Medium}, \emph{Large} is $2^{22}$, $2^{24}$ and $2^{26}$-vertex, respectively. For the matrix motif, we use 100, 1K and 10K two-dimensional matrix data with the same distribution and sparsity. For the transform motif, we use 16384, 32768 and 65536 two-dimension matrix data. For the other big data motifs, we use 1, 10 and 100 GB wikipedia text data, respectively.
For the AI motifs, we use three configurations in terms of input tensor sizes and channels. They are \emph{(224*224,64), (112*112,128) and (56*56,256)}. Among them, the first value indicates the dimension of input tensor, the second value indicates the channels, and  all of them use 128 as batch size. We choose these three configurations because they are widely used in neural network models~\cite{simonyan2014very}. Note that the dimension for all input tensors is 224 for \emph{Large} configuration, 112 for \emph{Medium} configuration and 56 for \emph{Small} configuration.
For the Pthread-version AI motifs, we use 1K, 10K, 100K images from ImageNet~\cite{deng2009imagenet}.
In the following subsections, we characterize the system and micro-architectural behaviors of data motifs with the \emph{Large} data size.
In Section~\ref{dataimpact}, we will analyze the impact of data input on characteristics with all data sizes.

\textbf{System and Micro-architecture Metrics}
We characterize the system and micro-architectural behaviors~\cite{van2016analytical} of the data motifs, which are significant for design and optimization~\cite{quinn2015using}.
For system evaluation, we report the metrics of CPU utilization, I/O Wait, disk I/O bandwidth, and network I/O bandwidth. The system metrics are collected through the proc file system.

For micro-architectural evaluation, we use the Top-Down analysis method~\cite{yasin2014top}, which categorizes the pipeline slots into four categories, including retiring, bad speculation, frontend bound and backend bound. Among them, retiring represents the useful work, which means the issued micro operations (uops) eventually get retired. Bad speculation represents the pipeline is blocked due to incorrect speculations. Frontend bound represents the stalls due to frontend, which undersupplies uops to the backend. Backend bound represents the stalls due to backend, which is a lack of required resources for new uops~\cite{pmutools}.
We use Perf~\cite{perftool}, a Linux profiling tool, to collect the hardware events referring to the Intel Developer\'s Manual~\cite{guide2011intel} and pmu-tools~\cite{pmutools}.

\subsection{System Evaluation}\label{system:exp}

Fig.~\ref{cpuiowait} presents the CPU utilization and I/O Wait of all data motifs. We find that Hadoop motifs have higher CPU utilization than Spark motifs, and suffer from less I/O Wait than Spark motifs do. Particularly, Hadoop motifs  take 80 percent CPU time.
The I/O Waits of AI data motifs are extremely lower than that of big data motifs.
For deep neural networks, even the total input data is large, the input layer loads a batch from disk every iteration, so data loading size from disk by the input layer occupies a very small proportion comparing to intermediate data, and thus introduces  little disk I/O requests.
Pthread motifs have less CPU utilization and I/O Wait in general, because Pthread motifs have less memory allocation and relocation operations than counterparts using other stacks. Moreover, the data loading time overlaps the processing time since computation is simple, except that Pthread Matmul has almost 100\% CPU utilization because of its high computation complexity and CPU-intensive characteristics.
TensorFlow motifs, such as AvgPool, Conv, Matmul, Maxpool, and Multiply, have taken most of CPU time, because these five motifs are CPU-intensive. Nevertheless, we also find that the other  AI motifs are not that CPU-intensive, such as Relu, Sigmoid, and Tanh. 

Fig~\ref{io} presents the network bandwidth and disk I/O bandwidth. For AI motifs, most
of them (e.g. matmul, relu, pooling, activation)  are executed in the hidden layers, and the intermediate states of hidden layers are stored in the memory. That is to say, the hidden layers consume the most resources of computation and memory storage, while the disk I/O for input layer is relatively minor. Our evaluation confirms this observation. Meanwhile, as mentioned in Section 4.1, we evaluate  both the TensorFlow and
Pthread implementations of AI motifs on one node for
apple-to-apple comparison. So we do not report the I/O behaviors of AI motifs.
We find that for all big data motifs, Spark stack has much larger network I/O pressure than that of Hadoop stack, because Spark stack has more data shuffles, so it needs transferring data from one node to another one frequently.
Five of the eight Spark implementations have smaller disk I/O pressure than that of Hadoop, because Spark targets in-memory computing.
Except Spark Matmul, Spark MD5 and Spark WordCount have larger disk I/O pressure than that of Hadoop counterparts.
Their disk I/O read sector numbers are nearly equal, while the write sector numbers are much larger.

\subsection{Micro-architecture Evaluation}

\begin{figure}[!t]
\centering
\includegraphics*[scale=0.4]{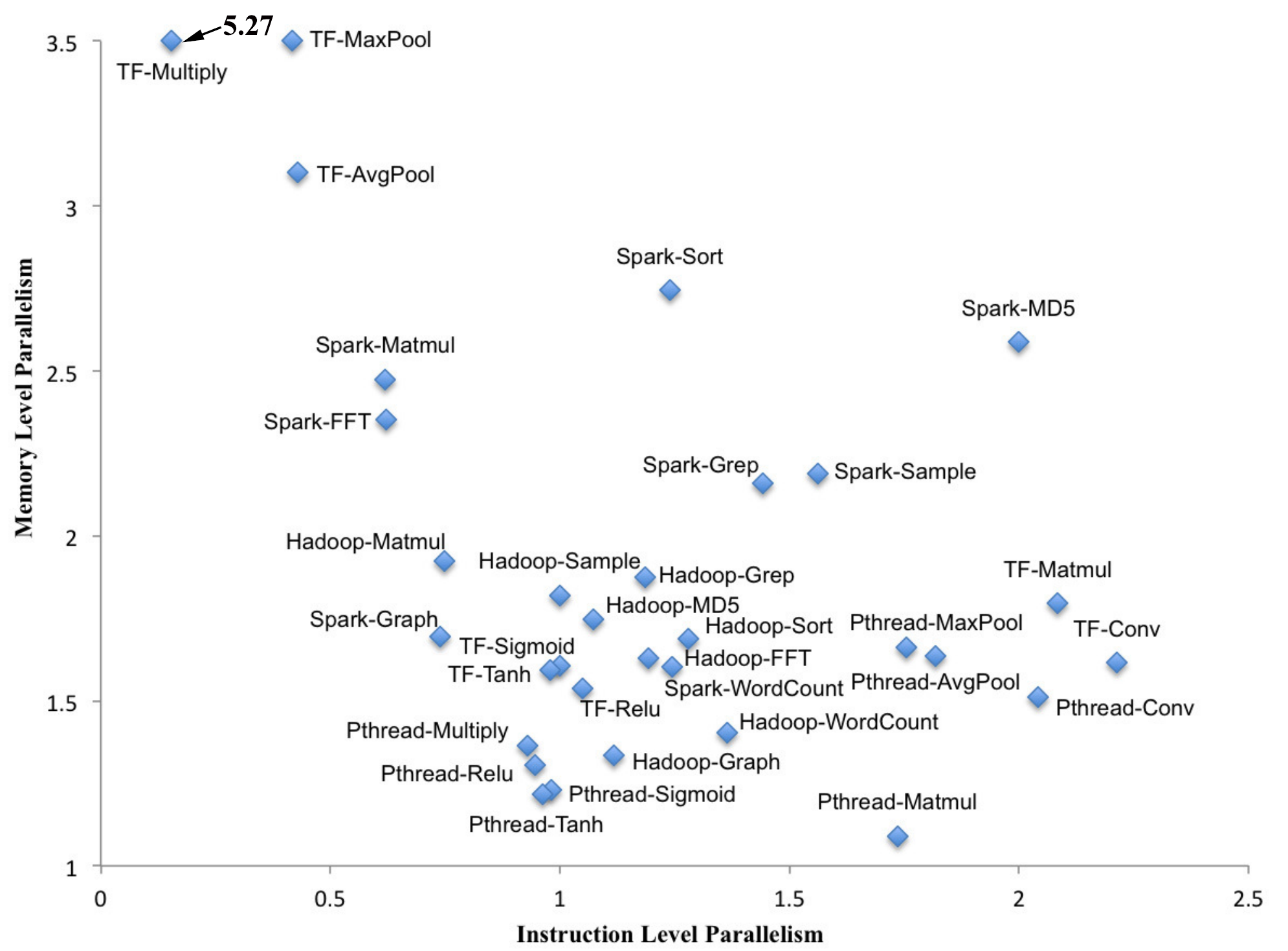}
\caption{Execution Performance of Data Motifs.} 
\label{ilpmlp}
\end{figure}

To better understand the data motifs, we analyze their performance and micro-architectural characteristics.

\textbf{Execution Performance} The execution performance indicates the overall running efficiency of the workloads~\cite{kim2016automatically}. We use the instruction level parallelism (ILP) and memory level parallelism (MLP) to reflect the execution performance. Among them, ILP measures the number of instructions that can be executed simultaneously. Here we use the retired instructions per cycle (IPC) to measure ILP. MLP indicates the parallelism degree that memory accesses can be generated and executed~\cite{glew1998mlp}. MLP is computed through dividing \emph{L1D\_PEND\_MISS.PENDING} by \emph{L1D\_PEND\_MISS.PENDING\_CYCLES}~\cite{pmutools}.
Fig.~\ref{ilpmlp} presents the ILP and MLP of all data motifs. We find that these motifs cover a wide range of ILP and MLP behaviors, reflecting distinct computation and memory access patterns.
For example, TensorFlow Multiply does element-wise multiplications and has high MLP (5.27) but extremely low ILP (0.15).  This is because that its computation is simple and has little data dependencies, so it generates many concurrent data loads, thus incurs a large amount of data cache misses. Also, max pooling and average pooling have high MLP.
The MLP of average pooling is lower than max pooling, because average computation involves many divide operations, and thus suffers from more stalls due to the delay of divider unit.
The software stack changes workload's computation and memory access patterns, which is also found in previous work~\cite{jia_bigDataBench_subset}. For example, both Hadoop FFT and Spark FFT are based on cooley-tukey algorithm~\cite{cooley1965algorithm}, while they have different parallelism degrees. Spark FFT is more memory-intensive and has higher MLP.

\begin{figure}[!t]
\centering
\includegraphics*[scale=0.53]{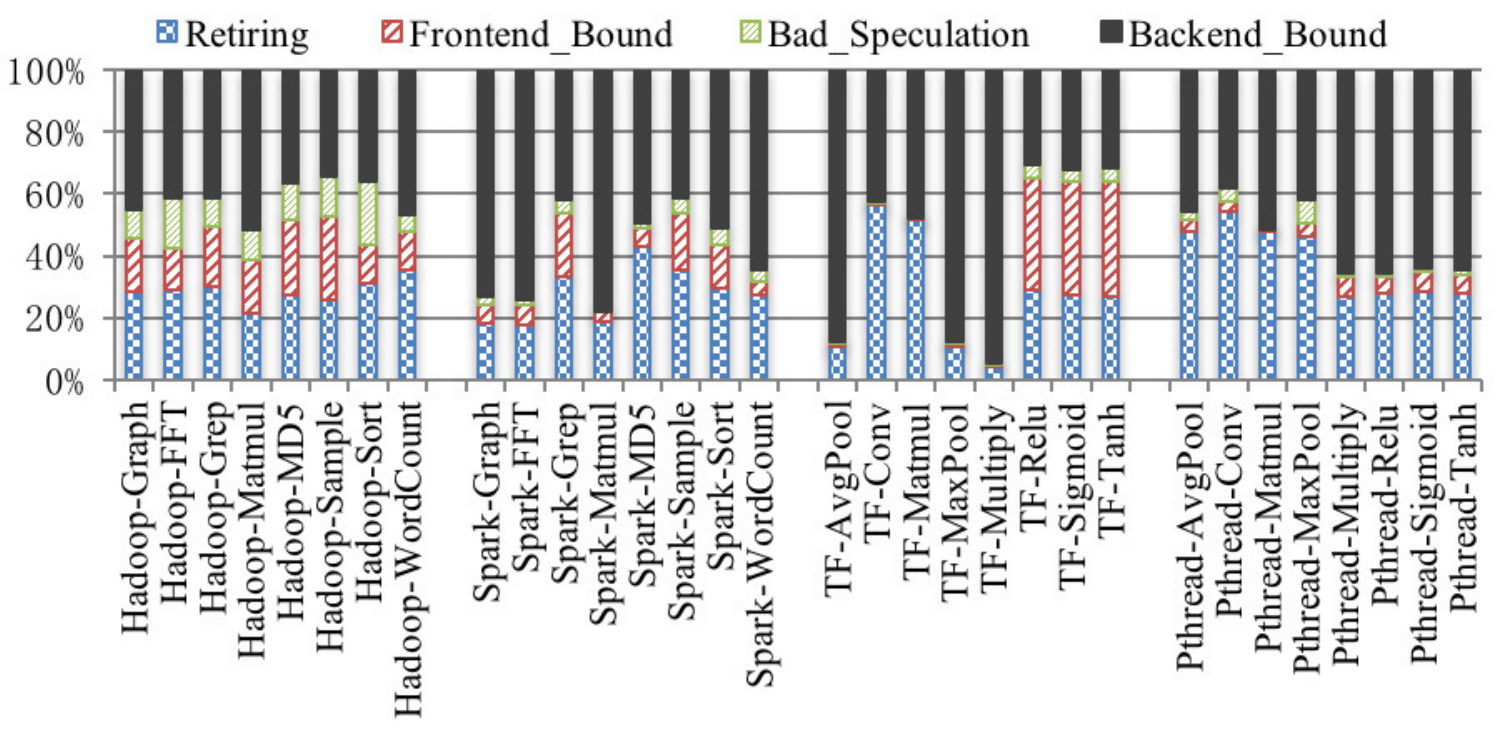}
\caption{The Uppermost Level Breakdown of Data Motifs.} 
\label{level1}
\end{figure}

\textbf{The Uppermost Level Breakdown} Fig.~\ref{level1} shows the uppermost level breakdown of all data motifs we evaluated. We find that these motifs have different pipeline bottlenecks. For Hadoop motifs, they suffer from notable stalls due to frontend bound and bad speculation.
Moreover, Hadoop motifs reflect nearly consistent bottlenecks, indicating the Hadoop stack impacts workload behaviors more than other stacks like Spark and TensorFlow.
For Spark motifs, which mainly compute in memory, they suffer from a higher percentage of backend bound than that of Hadoop counterparts. Spark Grep, Sample and Sort suffer from more frontend bound and their percentages of backend bound are smaller than the others.
The AI data motifs face different bottlenecks both on TensorFlow and Pthreads. Conv and Matmul have the highest IPC (about 2.2) and retiring percentages (about 50\% on TensorFlow). Max pooling, average pooling, and multiply have extremely low retiring percentages, which has been illustrated in above. However, activation operation like ReLU, sigmoid and tanh suffer from more frontend bound than backend bound.
For AI data motifs implemented with Pthread, their main bottleneck is backend bound. They suffer from little frontend and bad speculation stalls.

\begin{figure}[!t]
\centering
\includegraphics*[scale=0.52]{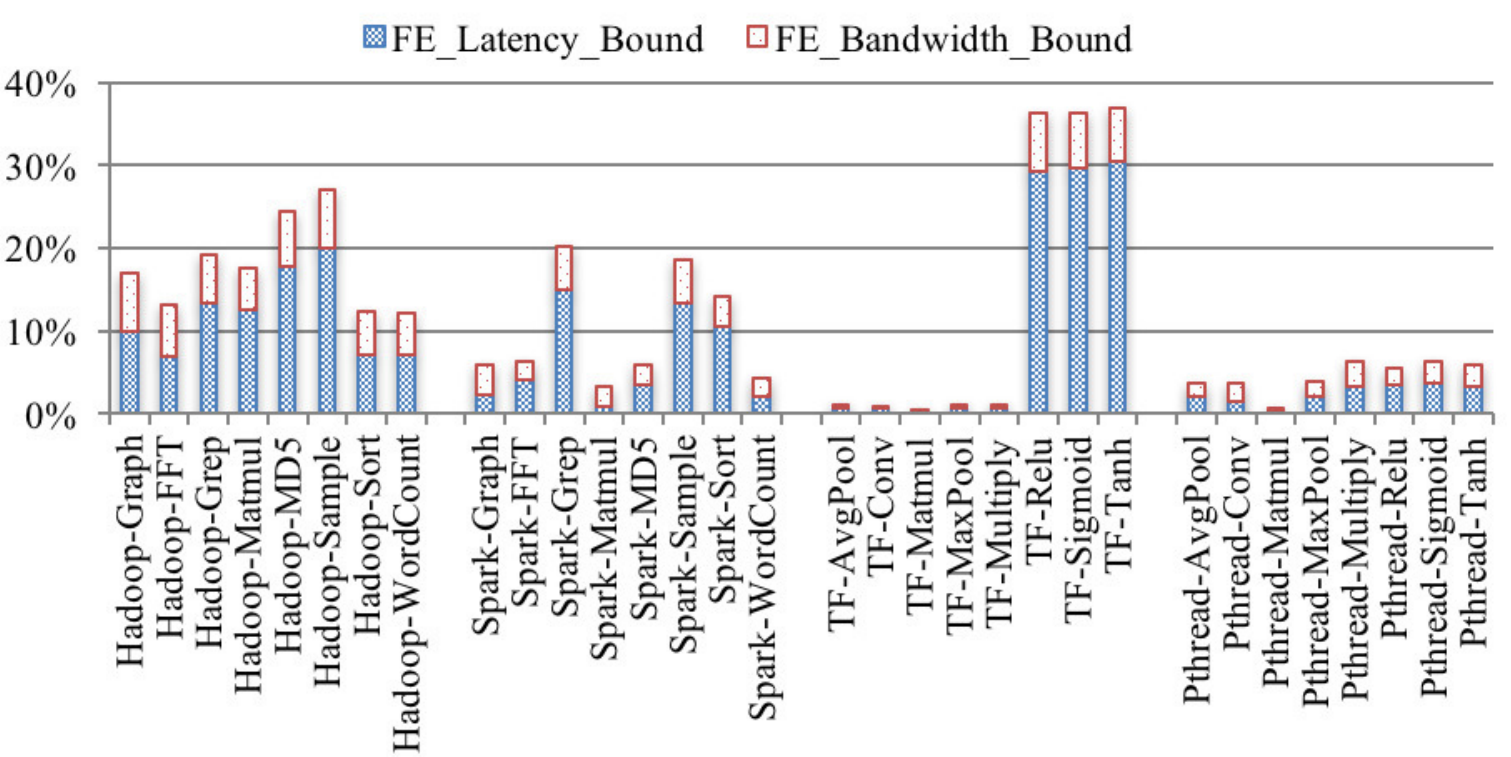}
\caption{The Frontend Breakdown of Data Motifs.} 
\label{l2front}
\end{figure}

\begin{figure}[!t]
\centering
\includegraphics*[scale=0.52]{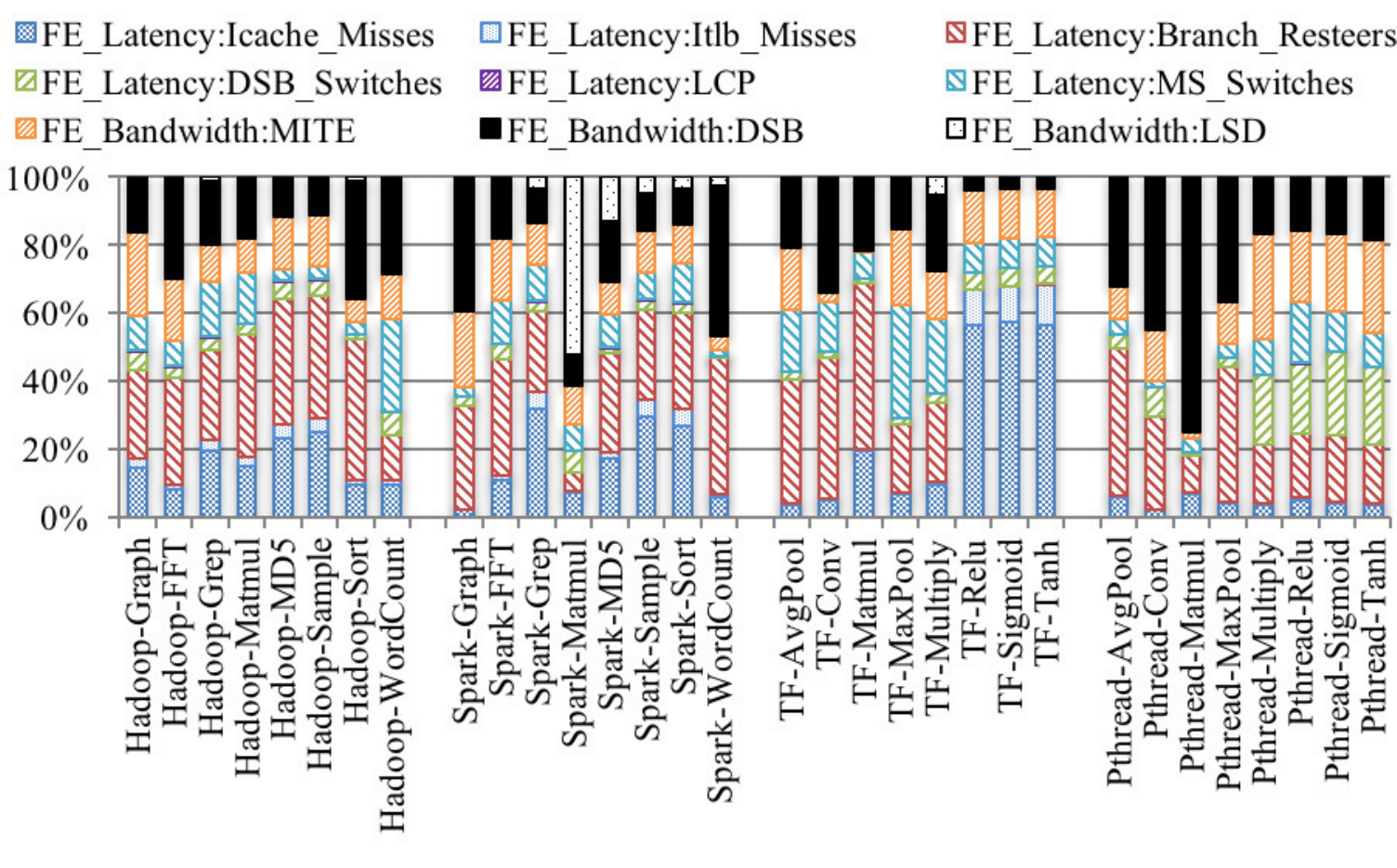}
\caption{The Frontend Latency Breakdown of Data Motifs.} 
\label{l3front}
\end{figure}

\textbf{Frontend Bound} Frontend bound can be split into frontend latency bound and frontend bandwidth bound. Among them, latency bound means the frontend delivers no uops to the backend, while bandwidth bound means delivering insufficient uops comparing to the theoretical value. Fig.~\ref{l2front} presents the frontend breakdown of the data motifs. We find that the main reason that incurs the frontend stalls is latency bound for almost all motifs that suffer from severe frontend bound.

We further investigate the reasons for the frontend latency bound and frontend bandwidth bound, respectively. Generally, the frontend latency bound are incurred by six reasons, including icache miss, itlb miss, branch resteers, DSB (Decoded Stream Buffer) switches, LCP (Length Changing Prefix), and MS (microcode sequencer) switches. Among them, icache miss and itlb miss are instruction cache miss and instruction tlb miss. Branch resteers means the delays to obtain the correct instructions, such as the delays due to branch misprediction. LCP measures the stalls when decoding the instructions with a length changing prefix.
Generally, uops comes from three places, including the decoded uops cache (DSB), legacy decode pipeline (MITE) and microcode sequencer (MS). DSB switches record the stalls caused by switching from the DSB to MITE. MS switches measure the penalty of switching to MS unit.
As for latency bandwidth bound, there are mainly two reasons: the inefficiency of MITE pipeline and the inefficient utilization of DSB cache. Additionally, LSD represents the stalls due to waiting the uops from the loop stream detector~\cite{lsd}.
Fig.~\ref{l3front} lists the latency and bandwidth bound breakdown of all data motifs.
For almost all data motifs, branch resteers is a main reason for the high percentage of frontend bound, except Spark Matmul and Relu, Sigmoid, Tanh on TensorFlow. For these three activation functions, nearly 60\% frontend bound is due to instruction cache miss.
On average, big data motifs implemented with Hadoop and Spark suffer from more icache misses than AI data motifs.
Moreover, MS switch is another significant factor that incurs frontend latency bound. Because big data and AI systems use many CISC instructions that cannot be decoded by default decoder, so they must be decoded by MS unit, and results in performance penalties.

\begin{figure}[!t]
\centering
\includegraphics[scale=0.52]{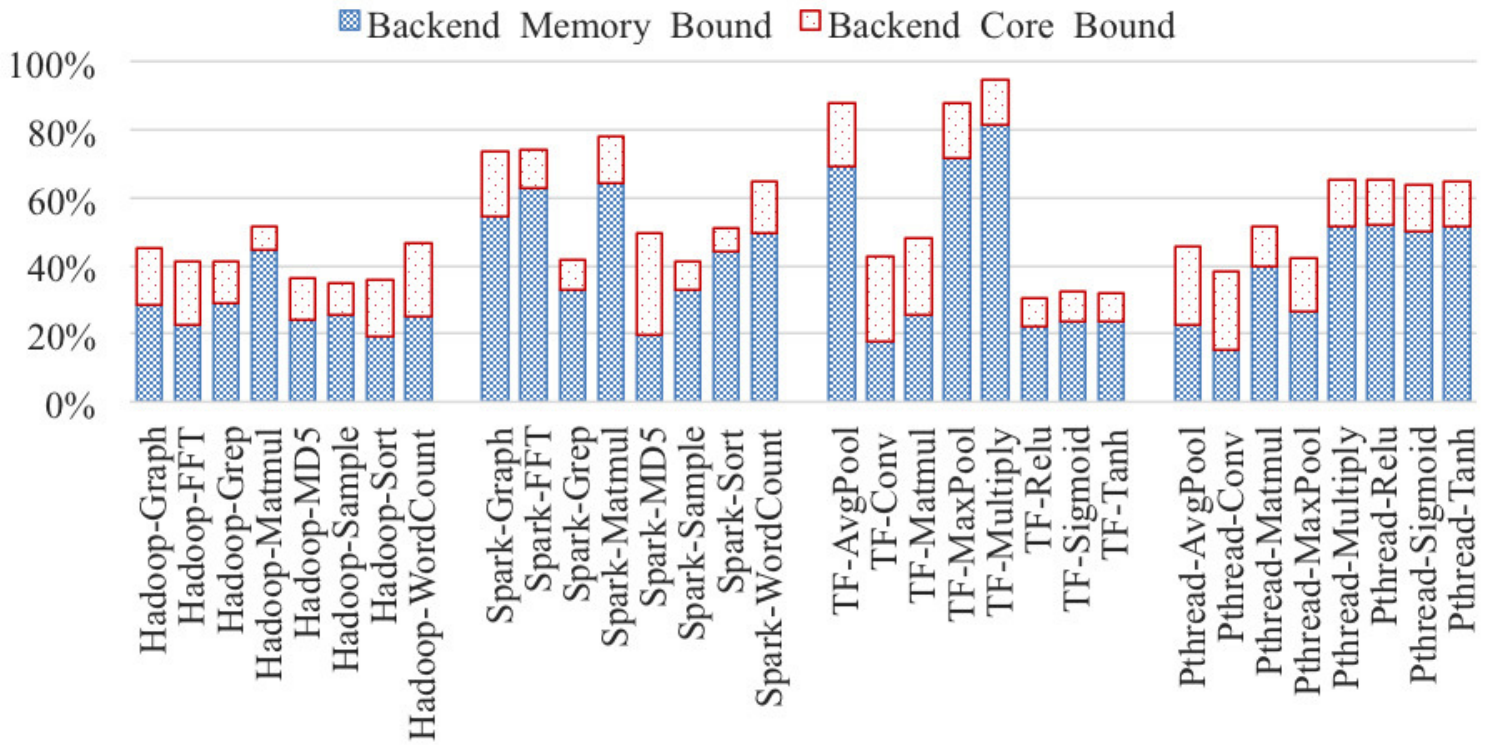}
\caption{The Backend Bound Breakdown of Data Motifs.} 
\label{l2back}
\end{figure}

\begin{figure}[!t]
\centering
\includegraphics[scale=0.52]{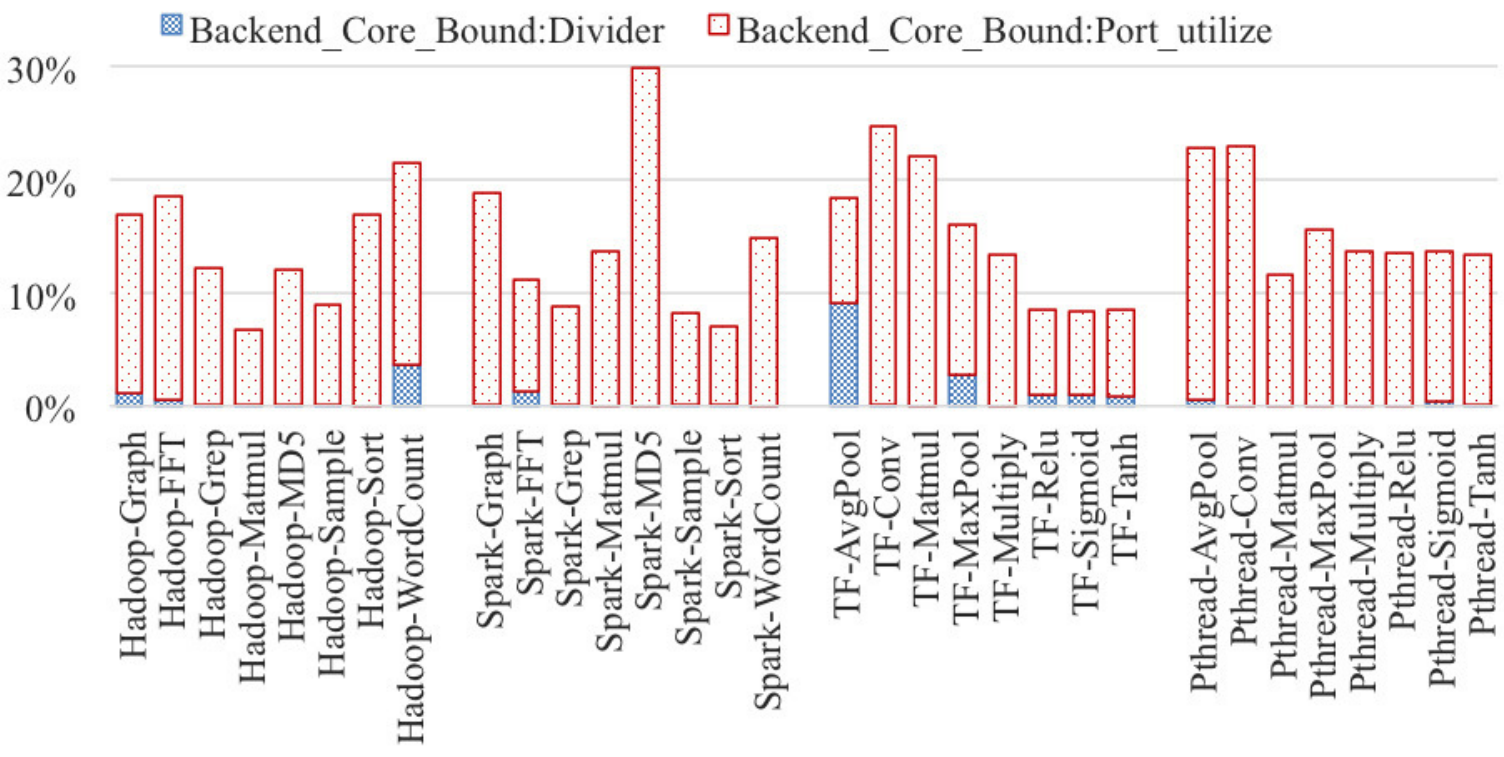}
\caption{The Backend Core Bound Breakdown of Data Motifs.} 
\label{l3core}
\end{figure}

\textbf{Backend Bound} Fig~\ref{l2back} presents the backend bound breakdown of data motifs, which are split into backend memory bound and backend core bound.
Backend memory bound is mainly caused by the data movement delays among different memory hierarchies.
Backend core bound is mainly caused by the lack of hardware resources (e.g. divider unit) or port under-utilization because of instruction dependencies and execution unit overloading.
We find that more than half of these data motifs suffer from more backend memory bound than core bound. However, for each software stack, there is at least one data motif that suffers from equal percentages of core bound or even more percentages of core bound than memory bound, such as Hadoop WordCount, Spark MD5, TensorFlow Conv and Pthread AvgPool.
Fig.~\ref{l3core} shows the core bound breakdown. We find that TensorFlow AvgPool and Hadoop WordCount suffer from significantly long latency of divider unit. While for Spark MD5 and TensorFlow Conv, which has the highest percentage of backend core bound, mainly suffer from the stalls due to port under-utilization. As for backend memory bound, we find that DRAM memory bound is much severe than level 1, 2, and 3 cache bound for almost all big data and AI motifs, indicating that the memory wall~\cite{wulf1995hitting} still exists and needs to be optimized.

\begin{figure*}[!t]
\centering
\includegraphics*[scale=0.38]{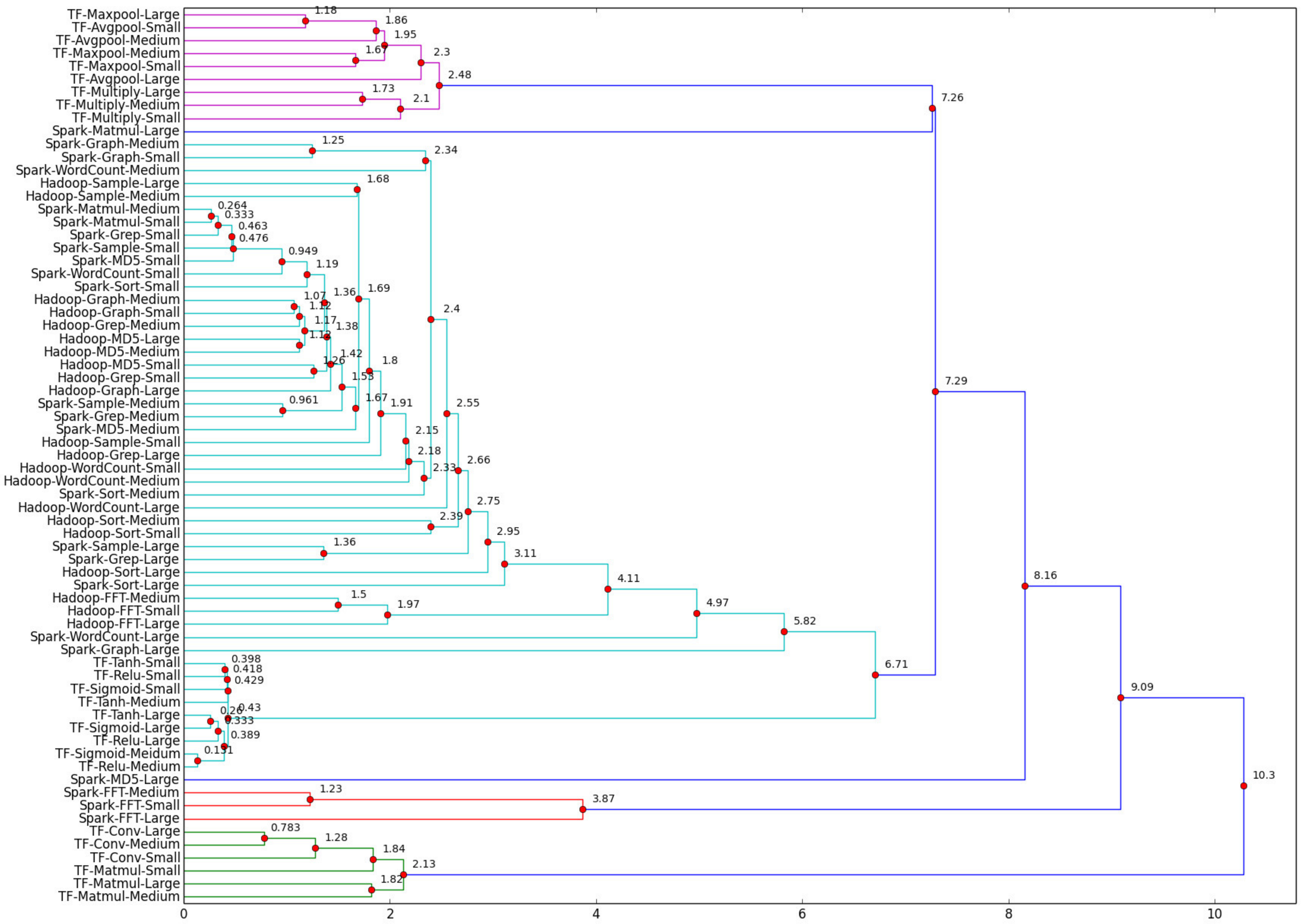}
\caption{Linkage Distance of Data Motifs.} 
\label{similar}
\end{figure*}

\section{Impact of Data Input}\label{dataimpact}

In this section, we evaluate the impact of data input on system and micro-architecture behaviors from the perspectives of size, source, type, and pattern. For type and pattern evaluation, we use Sort and FFT as an example, respectively.

\subsection{Impact of Data Size}

Based on all sixty metrics spanning system and micro-architecture we evaluated in  Section~\ref{evaluation}, we conduct a coarse-grained similarity analysis using PCA (Principal Component Analysis)~\cite{jolliffe1986principal} and hierarchical clustering~\cite{johnson1967hierarchical} methods on three data size configurations. Fig.~\ref{similar} presents the linkage distance of all data motifs, which indicates the similarity of system and micro-architecture behaviors. Note that the smaller the linkage distance, the more similar the behaviors.
We find that data motifs with small data size are more likely to be clustered together. A small data size will not fully utilize the system and hardware resources, hence that they tend to reflect similar behaviors. However, for the motif that  is computation intensive and has high computation complexity, even with the large data set, it will be clustered together with small data set. For example, FFTs with three data size configurations are clustered together for both Hadoop and Spark version. AI Motifs with TensorFlow implementations are also greatly affected by the input data size. However, they reflect distinct behaviors with big data motifs implemented with Hadoop and Spark, with the least linkage distance of 6.71.

\begin{figure}[!t]
\centering
\includegraphics*[scale=0.47]{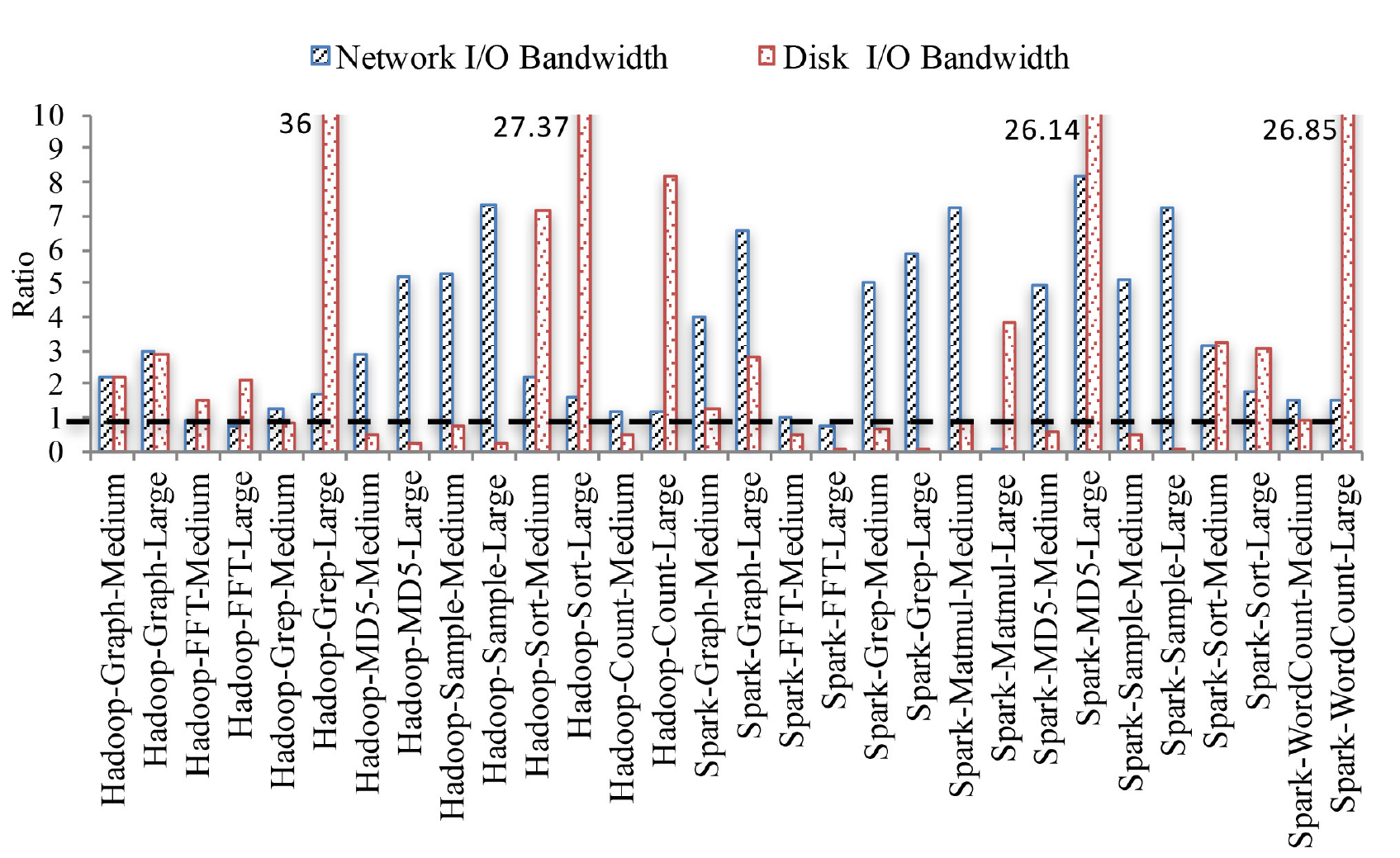}
\caption{Impact of Data Size on I/O Behaviors.} 
\label{ioimpact}
\end{figure}

\textbf{Impact of Data Size on I/O Behaviors}
We evaluate the impact of data size on I/O behaviors using the fully distributed Hadoop and Spark motif implementations. 
Using the I/O bandwidth of \emph{Small} data size as baseline, we normalize the I/O bandwidth of \emph{Medium} and \emph{Large} data size, as illustrated in Fig.~\ref{ioimpact}. The bold black horizontal line in Fig.~\ref{ioimpact} shows the equal line with the small input. That is to say, the value higher than the line means larger I/O bandwidth than the value of the small input. Here we do not report the performance data of the AI motifs because the disk I/O behavior is little in neural network modelling,  which we have illustrated in Subsection~\ref{system:exp}.
We find that almost for all data motifs, their I/O behaviors are sensitive to the data size. When the data size large enough, the whole data can not be stored in  memory, then the data have to be swapped in and swapped out frequently, and hence put great pressure on disk I/O access. Modern big data and AI systems adopt a distributed manner, with the data storing on an distributed file system, such as HDFS~\cite{shvachko2010hadoop}, the data shuffling or data unbalance will generate a large amount of network I/O.

\begin{figure*}[!t]
\centering
\includegraphics*[scale=0.53]{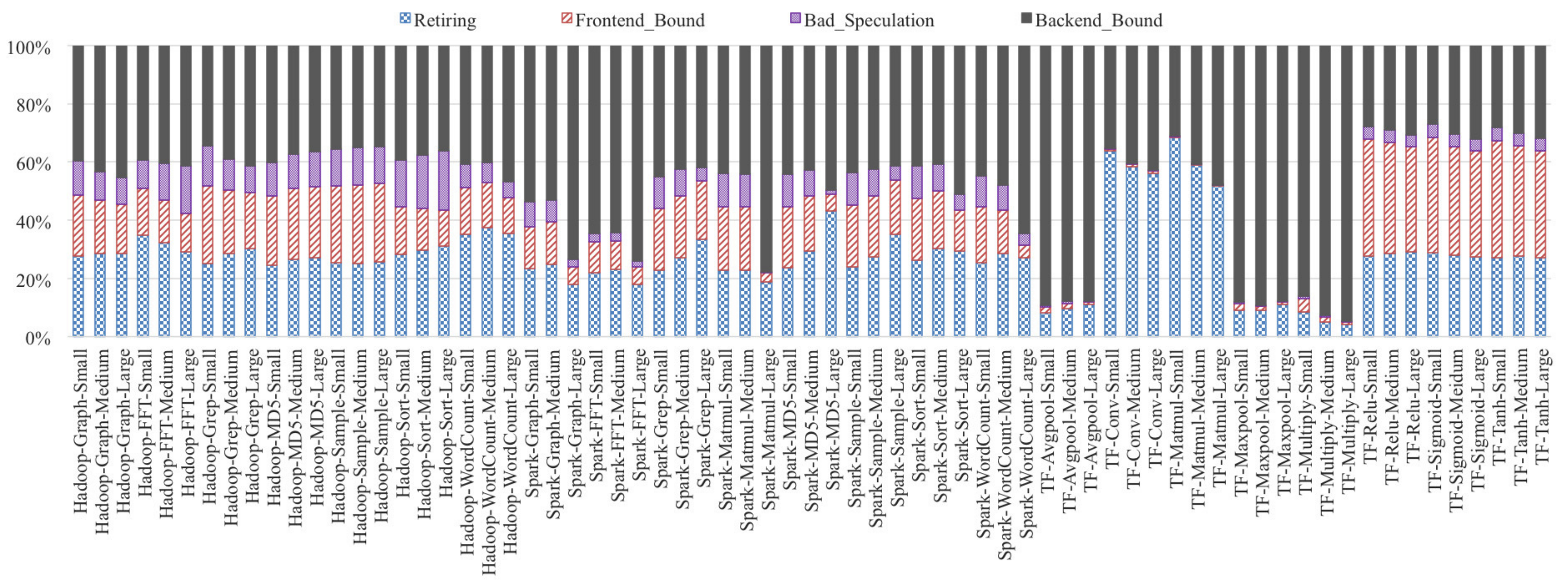}
\caption{Impact of Data Size on Pipeline Efficiency.} 
\label{pipimpact}
\end{figure*}

\textbf{Impact of Data Size on Pipeline Efficiency}
We further measure the impact of data size on pipeline efficiency. As shown in Fig.~\ref{pipimpact}, we find that with the data size increases, the percentage of frontend bound decrease, while the percentage of backend bound increase. For example, Spark Matmul with large input size decrease nearly 20\% of frontend bound and increase more than 30\% of backend bound. As the data size increase, the high-speed cache and even memory are unable to hold all of them, and further incur many data cache misses, resulting in large penalties due to memory hierarchy.

\subsection{Impact of Data Pattern}

\begin{figure*}
\centering
\subfigure[System Behavior with Different Patterns.]{
\label{figP:subfig:a} 
\includegraphics[scale=0.5]{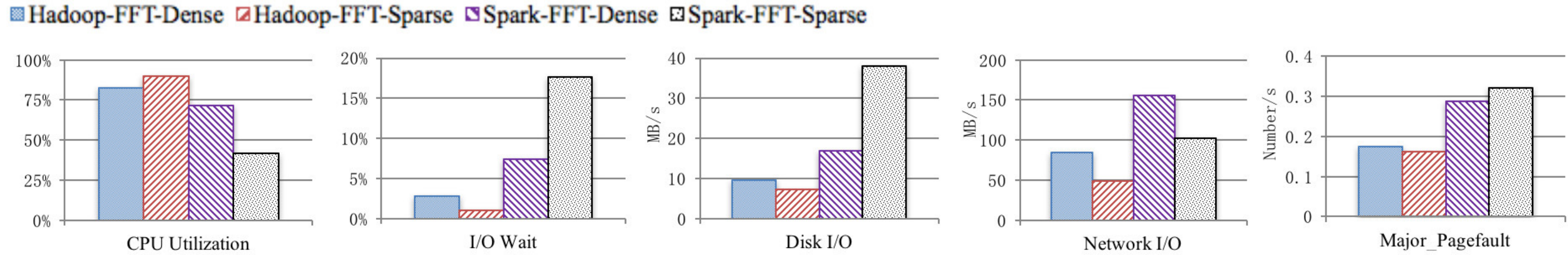}}
\hspace{1in}
\subfigure[Micro-architecture Behavior with Different Patterns.]{
\label{figP:subfig:b} 
\includegraphics[scale=0.5]{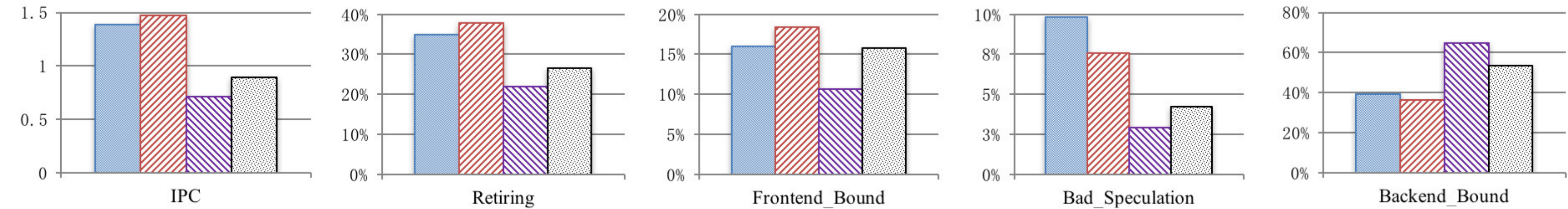}}
\caption{Impact of Data Pattern on Data Motifs.}
\label{figP:subfig} 
\end{figure*}

Data pattern and data distribution impact the workload performance significantly~\cite{xie2018cvr,yilmaz2016autotuning}. To evaluate the impact of data pattern on the motifs, we use two different patterns of dense matrix and sparse matrix, to run FFT motif as an example. The matrix sparsity indicates the ratio of zero value among all matrix elements. With different sparsity, the data access patterns vary, and thus reflect different behaviors.

We use two 16384$\times$16384 matrixes as the input for the FFT motif, with the one having 10\% sparsity and the other one 90\% sparsity.
Fig.~\ref{figP:subfig} shows the impact of data pattern on the data motifs from system (Fig.~\ref{figP:subfig:a}) and micro-architecture perspectives(Fig.~\ref{figP:subfig:b}). We find that using the matrix with high sparsity, the network I/O and disk I/O are nearly half of the values using the dense matrix, and the major page fault per second is almost the same. Spark motifs suffer from more I/O pressure than Hadoop motifs. As for pipeline bottlenecks, sparse data input incurs more frontend stalls while less backend stalls.

\subsection{Impact of Data Type and Source}

\begin{figure*}
\centering
\subfigure[System Behavior with Different Types.]{
\label{fig:subfig:a} 
\includegraphics[scale=0.5]{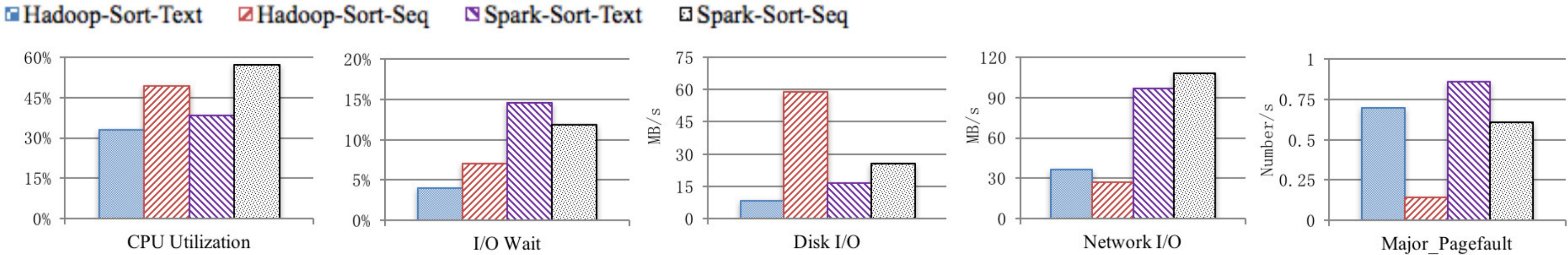}}
\subfigure[Micro-architecture Behavior with Different Types.]{
\label{fig:subfig:b} 
\includegraphics[scale=0.5]{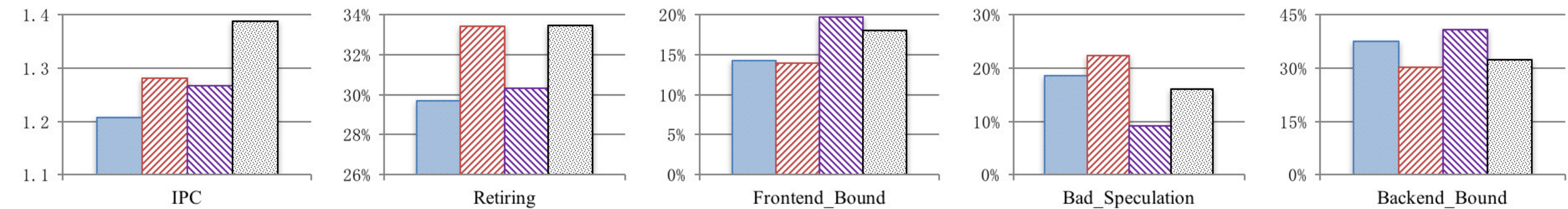}}
\caption{Impact of Data Type and Source on Data Motifs.}
\label{fig:subfig} 
\end{figure*}

Data types and sources are of great significance for read and write efficiency~\cite{eeckhout2003quantifying}, considering their storage format and targeted scenarios, such as the supports for splitable files and compression level.
To evaluate the impact of the data type and source on system and micro-architecture behaviors, we use two different data types for Sort motif, with the same data size of 10 GB. Two types are un-structured wikipedia text data and semi-structured sequence data. Wikipedia text file is laid out in lines and each line records an article content. Sequence files are flat files that consist of key and value pairs, stored in binary format. Fig.~\ref{fig:subfig} lists the impact of data type on data motifs from the system (Fig.~\ref{fig:subfig:a}) and micro-architecture aspects (Fig.~\ref{fig:subfig:b}). We find that the difference between using text type and sequence type ranges from 1.12 times to 7.29 times from the system aspects. 
Using text data type, the CPU utilization is lower than using sequence data, which indicates that using sequence data has better performance.
Moreover, both Hadoop Sort and Spark Sort suffer from more major page faults and further impact the execution performance, because of page loads from disk. Note that we use the major page fault number per second in Fig.~\ref{fig:subfig} and the total number during the running process is about 100 to 200. 
Even with the same amount of data size, their network I/O and disk I/O bandwidth still have a great difference. We find that the sequence format have larger requirements for I/O bandwidth than the text format.
From the micro-architecture aspect (Fig.~\ref{fig:subfig:b}), Sort with different data types reflect different percentages of pipeline bottlenecks. With the text format, backend bound bottleneck is more severe, especially backend memory bound, which indicates that they waste more cycles to wait for the data from cache or memory.

\section{Related Work}

Our big data and AI motifs are inspired by previous successful abstractions in other application scenarios.
The \emph{set} concept in relational algebra~\cite{codd1970relational} abstracted five primitive and fundamental operators, setting off a wave of relational database research. The set abstraction is the basis of relational algebra and theoretical foundation of database.
Phil Colella~\cite{colella2004defining} identified seven motifs of numerical methods which he thought would be important for the next decade. Based on that, a multidisciplinary group of Berkeley researchers proposed 13 motifs which were highly abstractions of parallel computing, capturing the computation and communication patterns of a great mass of applications~\cite{asanovic2006landscape}.
National Research Council proposed seven major tasks in massive data analysis~\cite{council2013frontiers}, which they called giants. These seven giants are macroscopical definition of problems in massive data analysis from the perspective of mathematics, while our eight classes of motifs are main time-consuming units of computation in the Big Data and AI workloads.

Application kernels~\cite{bailey1991parallel,dongarra2003linpack} also aim at scaling down the run time of the real applications, while preserving the main characteristics of the workload.
Consisting of the major function of the application, Kernel tries to cover the bottleneck of the real application.
But kernel is still hard to understand the complex and diversity big data and AI workloads~\cite{bailey1991parallel,lilja2005measuring}.
Other than that, kernel mainly focuses on the CPU and memory behaviors, and pays little attention to the I/O, which is also important for many real applications, especially in an era of data explosion.

\section{Conclusions}

In this paper, we answer what are abstractions of time-consuming units of computation
in big data and AI workloads.
We identify eight data motifs among a wide variety of big data and AI workloads, including Matrix, Sampling, Logic, Transform, Set, Graph, Sort and Statistic computations. 
We found the combinations of one or more data motifs with different weights in terms of runtime can
describe most of big data and AI workloads we investigated~\cite{gao2018proxy}.
We implement the data
motifs for big data and AI separately, including the big data motif
implementations using Hadoop, Spark, Pthreads, and the AI data
motif implementations using TensorFlow, Pthreads, considering
the impact of data type, data source, data size, and data pattern.
 We release them as the micro benchmarks of an open-source Big Data and AI benchmark suite---BigDataBench, publicly available from \url{http://prof.ict.ac.cn/BigDataBench}. 
From the system and micro-architecture perspectives, we comprehensively characterize the behaviors of data motifs
 and identify their bottlenecks. Further, we measure the impact of data type, data source, data pattern and data size on their behaviors.
We find that these data motifs cover a wide variety of performance space, from the perspectives of system and micro-architecture behaviors.
Moreover, the behavior of each data motif is not only influenced by its algorithm, but also largely affected by the type, source, size, and pattern of input data.
 We believe our work is an important step toward not only Big Data and AI benchmarking, but also domain-specific hardware and software co-design.

\section{Acknowledgements}

This work is supported by the National Key Research and Development Plan of China (Grant No. 2016YFB1000600 and 2016YFB1000601). The authors are very grateful to anonymous reviewers for their insightful feedback and Dr. Zhen Jia for his valuable suggestions.


%

\appendix
\section{Artifact appendix}


\subsection{Abstract}

{\em The artifact contains our big data and AI motif implementations on Hadoop, Spark, Pthreads, and TensorFlow stacks. It can support the characterization results in Chapter four and Chapter five of our PACT 2018 paper \textbf{Data Motifs: A Lens Towards Fully Understanding Big Data and AI Workloads}. To validate the results, deploy the experiment environment and profile the benchmarks.}

\subsection{Artifact check-list (meta-information)}


{\small
\begin{itemize}
  \item {\bf Program: Data motif implementations}
  \item {\bf Compilation: GCC 4.8.5; Python 2.7.5; Java 1.8.0\_65}
  \item {\bf Data set: generated by BigDataBench}
  \item {\bf Run-time environment: CentOS 7.2, Linux Kernel 4.1.13 with Perf tool}
  \item {\bf Hardware: Processor supporting Top-Down analysis, above Sandy Bridge series, and the performance events corresponding to the processor}
  \item {\bf Run-time state: Disable Hyper-Threading}
  \item {\bf Execution: root user or users that can execute sudo without password}
  \item {\bf Output: the system and micro-architecture profiling results}
  \item {\bf Experiment: Deploy the data motifs and corresponding software stacks; run benchmarks; profile using perf; output the results}
  \item {\bf Workflow frameworks used? No}
  \item {\bf Publicly available?: Yes}
\end{itemize}

%

\subsection{Description}

\subsubsection{How delivered}

%

The data motifs are the micro benchmarks of BigDataBench 4.0---an open source big data and AI benchmark suite. Download link:

http://prof.ict.ac.cn/bdb\_uploads/bdb\_4/pact2018.tar.gz

All the related files are under the \textbf{"pact2018"} directory, please refer to README for detailed description.
Note that to obtain accurate performance data, the user should make sure there is no other motif running before run a motif. The running scripts we provide suit for our cluster environment, like the node ip/hostname and port number, if you download and use it in your cluster environment, you need to modify the scripts to suit for your environment.

\subsubsection{Hardware dependencies}

The data motifs can be run on all processors that can deploy Hadoop, Spark, TensorFlow and Pthread stacks. However, for Top-Down analysis, due to the performance counter limitations, we suggest the Intel Xeon processors, above Sandy Bridge series. Also, user need to find the performance counters corresponding to specific processor. We have provided profiling scripts for Xeon E5-2620 V3 (Haswell) processor.

\subsubsection{Software dependencies}

JDK 1.8.0\_65; Hadoop 2.7.1; Spark 1.5.2; TensorFlow 1.0; GCC 4.8.5.

\subsubsection{Data sets}

We provide data generators for text, sequence, graph, and matrix data. Users can find the data generation method in the README file or BigDataBench user manual. The generation parameter used in our paper for the graph motif is 22 (Small), 24 (Medium), 26 (Large), respectively.

\subsection{Installation}

{\em User need to install Hadoop, Spark, GCC and TensorFlow. The install details can be found in the User Manual of BigDataBench. We provide "Makefile" for pthread motifs. For all data motifs, we provide running scripts in our package.}

\subsection{Experiment workflow}

Before profiling system and micro-architecture metrics of one motif, users should make sure there is no other motif/workload running.

\subsubsection{Data generation}

We provide text, graph, matrix, and sequence data generators under data-generator directory.
To generate large, medium, small data used in our paper, we provide a script "data-generator.sh". Make sure hadoop is running, because the script upload the generated data to HDFS. The script running command:

\#sh data-generator.sh <format> <datasize>

Note that <format> can be text, seq, graph or matrix, and <datasize> can be large, medium or small.

Also, the generators support generate other data size the user needed.

\textbf{Graph data generation:}

\#cd \$pact2018/data-generator/genGraphData

\#./genGraph.sh <log2\_vertex>

For example, ./genGraph.sh 26 for 2\^26-vertex graph data.

\textbf{Matrix data generation:}

\#cd \$pact2018/data-generator/genMatrixData

For floating-point data: \#./generate-matrix.sh <row\_num> <colum\_num> <sparsity>

For integer data: \#./generate-matrix-int <row\_num> <colum\_num> <sparsity>

The sparsity means "sparsity" percentage elements are zero.

\textbf{Text data generation:}

\#cd \$pact2018/data-generator/genTextData

\#./genText.sh <size>

Note that the parameter size means "size" gigabytes text data.

\textbf{Sequence data generation:}

Transfer the wiki text data to sequence data, so user should generate text data first and put it on HDFS, for example, "wiki-10G" data are on HDFS.

\#cd \$pact2018/data-generator/genSeqData

\#./sort-transfer.sh <size>

\subsubsection{Run the workloads.}

We provide running scripts for all workloads. During the running process, the profiling scripts are started to sample the system and architecture metrics.

\textbf{For Hadoop motifs:}

1) Under pact2018 directory

2) Start Hadoop: \#./start-hadoop.sh

3) Choose one Hadoop motif: \#./run-hadoop.sh motif datasize

Note that datasize parameter can be "large", "medium" or "small", means using large/medium/small data size,respectively.
For example: \#./run-hadoop.sh graph large

\textbf{For Spark motifs:}

1) Under pact2018 directory

2) Start Spark: \#./start-spark.sh

3) Choose one Spark motif: \#./run-spark.sh motif datasize

Note that datasize parameter can be "large", "medium" or "small", means using large/medium/small data size,respectively.
For example: \#./run-spark.sh graph large

\textbf{For TensorFlow motifs:}

1) Under pact2018 directory

2) Choose one TensorFlow motif: \#./run-tensorflow.sh motif datasize

Note that datasize parameter can be "large", "medium" or "small", means using large/medium/small data size,respectively.
For example: \#./run-tensorflow.sh relu large

\textbf{For Pthread motifs:}

1) Under pact2018 directory

2) Choose one Pthread motif: \#./run-pthread.sh motif datasize

Note that datasize parameter can be "large", "medium" or "small", means using large/medium/small data size,respectively.
For example: \#./run-pthread.sh relu large

The sampling results of system and micro-architecture metrics are under "result" directory. We provide processing scripts for computing the result and plot the figures. Please refer to "README" file for the details.

\subsubsection{Process the metric data and plot the figures}

We provide processing scripts and figure plotting scripts to generate the figures used in the paper. Note that the sampling results are saved under "result" directory when test finished.

1) Compute the performance data and save them in an excel file.

\#python lsdata.py result result\_new 1

Parameter "result" means the input directory which contains the sampling results; Parameter "result\_new" means the output excel file name and the output file is result\_new.xls.

2) Plot the figures and save them as png image format

\#python plot.py result\_new.xls

Parameter "result\_new.xls" is the excel file generated by the first step. After running the command, several png files will be generated. In addition, "pact-AE.txt" is generated for linkage distance analysis.

3) Linkage distance computing

\#\$pact2018/Linkage-Distance

\#python hiclust\_wiht\_newpca.py pact-AE.txt

Parameter "pact-AE.txt" is the text file generated by the second step. After running the command, a png file will be generated under the Linkage-Distance directory, which is used as Figure 12 in our paper.

\subsection{Evaluation and expected result}

{\em To evaluate the system and micro-architecture performance of data motifs, users need to run those motifs and profile them.
These data motifs should reflect similar characteristics like figures in Chapter 4 and Chapter 5. Our profiling scripts sample the performance data every 1 second during the whole motif runtime, and the performance data possibly vary within a slightly variation for each run.}

\subsection{Experiment customization}

Users can run these data motifs for different benchmarking purpose, e.g. software stack comparison, different aspects of system and architecture characterizations. Also, the data motifs can be deployed on different processors and cluster scales.

\subsection{Notes}

For the artifact evaluation, since every motifs need to run three times for collecting dozens of performance events, it may cost several weeks to profiling all motifs, which is too expensive for the artifact evaluation. So we provide the profiling scripts and the profiling data used in our paper, which are suit for our Haswell processor configurations. Since the platform configurations of software (e.g. Hadoop/Spark configuration) and hardware (e.g. memory capacity, BIOS configuration) may be different, so the performance data may be different on another platform.


\end{document}